\documentclass[11pt,a4paper]{article}
\usepackage{jheppub}
\usepackage{graphicx}
\usepackage{amssymb}
\usepackage{amsmath}
\usepackage{epsfig}
\usepackage {longtable}
\usepackage {soul}
\usepackage{bm}
\usepackage{slashed}
\usepackage{url}
\usepackage{multirow}
\usepackage{bigstrut}
\usepackage{wrapfig}

\newcommand{\beq}{\begin{equation}}
\newcommand{\eeq}{\end{equation}}
\def\bsp#1\esp{\begin{split}#1\end{split}}
\def\bal#1\eal{\begin{align}#1\end{align}}
\newcommand{\beeq}{\begin{eqnarray}}
\newcommand{\eeeq}{\end{eqnarray}}

\newcommand{\lo}       {{\rm LO}}
\newcommand{\nlo}      {{\rm NLO}}

\newcommand{\lhe}      {{\rm LHE}}
\newcommand{\ckm}      {{\rm CKM}}

\newcommand{\pdf}      {{\rm PDF}}
\newcommand{\pdg}      {{\rm PDG}}
\newcommand{\pythia}   {\texttt{PYTHIA}}
\newcommand{\mcfm}     {\texttt{MCFM}}
\newcommand{\fastjet}     {\texttt{Fastjet 3.0.0}}
\newcommand{\py}       {\texttt{PY}}
\newcommand{\herwig}   {\texttt{HERWIG}}
\newcommand{\hw}       {\texttt{HW}}
\newcommand{\powhel}   {\texttt{PowHel}}

\newcommand{\powhegbox}{\texttt{POWHEG-BOX}}

\newcommand{\helacnlo} {\texttt{HELAC-NLO}}

\newcommand{\helaconeloop} {\texttt{HELAC-1LOOP}}

\newcommand{\helacphegas} {\texttt{HELAC-Phegas}}

\newcommand{\lhapdf}           {\texttt{LHAPDF}}

\newcommand{\as}       {\ensuremath{\alpha_{\mathrm{s}}}}

\newcommand{\pt}       {\ensuremath{p_\bot}}

\newcommand{\pTmiss}{\ensuremath{\slash\hspace*{-5pt}{p}_{\perp}}}

\newcommand{\ptt}      {\ensuremath{p_{\bot,\,\mathrm{t}}}}
\newcommand{\ptttbar}  {\ensuremath{p_{\bot,\,\mathrm{t\,\bar{t}}}}}

\newcommand{\qq}       {\ensuremath{\mathrm{q}}}

\newcommand{\qaqp}     {\ensuremath{\mathrm{\bar{q}'}}}

\newcommand{\tq}       {\ensuremath{\mathrm{t}}}
\newcommand{\taq}      {\ensuremath{\mathrm{\bar{t}}}}

\newcommand{\tT}       {\ensuremath{\mathrm{t\,\bar{t}}}}

\newcommand{\mur}      {\ensuremath{\mu_\mathrm{R}}}
\newcommand{\muf}      {\ensuremath{\mu_\mathrm{F}}}

\newcommand{\tev}      {\ensuremath{\mathrm{TeV}}}
\newcommand{\gev}      {\ensuremath{\mathrm{GeV}}}

\newcommand{\mt}       {\ensuremath{m_{\mathrm{t}}}}
\newcommand{\mz}       {\ensuremath{m_Z}}
\newcommand{\mw}       {\ensuremath{m_W}}

\newcommand{\lomath}    {\ensuremath{\mathrm{LO}}}
\newcommand{\nlomath}   {\ensuremath{\mathrm{NLO}}}

\newcommand\figs[2]    {Figs.\,{\ref{#1}} and ~\ref{#2}}

\newcommand\sect[1]    {Sect.\,{\ref{#1}}}

\newcommand{\ttv}     {\ensuremath{\mathrm{t\ \bar{t}}\  V}}
\newcommand{\ttw}     {\ensuremath{\mathrm{t\ \bar{t}}\  W^\pm}}
\newcommand{\ttwp}    {\ensuremath{\mathrm{t\ \bar{t}}\ W^+}}
\newcommand{\ttwm}    {\ensuremath{\mathrm{t\ \bar{t}}\ W^-}}
\newcommand{\ttz}     {\ensuremath{\mathrm{t\ \bar{t}}\ }Z}
\newcommand{\tth}     {\ensuremath{\mathrm{t\ \bar{t}}\ }H}
\newcommand{\ttj}     {\ensuremath{\mathrm{t\ \bar{t}}\ j}}

\title{\ttw\ and \ttz\ hadroproduction at NLO accuracy in QCD with Parton
Shower and Hadronization effects}
\author[a]{M. V. Garzelli,}
\author[b]{A. Kardos,}
\author[c,d]{C. G. Papadopoulos,}
\author[b]{Z. Tr\'ocs\'anyi}
\affiliation[a]{Laboratory for Astroparticle Physics, University of Nova
Gorica,\\ SI-5000 Nova Gorica, Slovenia }
\affiliation[b]{Institute of Physics and MTA-DE Particle Physics Research
Group, University of Debrecen,\\
H-4010 Debrecen P.O.Box 105, Hungary}
\affiliation[c]{Institute of Nuclear Physics, NCSR 
$\Delta\eta\mu${\it \'o}$\kappa\rho\iota\tau o \varsigma$, \\
GR-15310 Athens, Greece}
\affiliation[d]{CERN, TH-Unit, CH-1211 Geneva 23, Switzerland}

\emailAdd{kardos.adam@science.unideb.hu}
\emailAdd{garzelli@mi.infn.it}
\emailAdd{costas.papadopoulos@cern.ch}
\emailAdd{Zoltan.Trocsanyi@cern.ch}

\abstract{
We present theoretical predictions for the hadroproduction of \ttwp,
\ttwm\ and \ttz\ at LHC  as obtained by matching numerical computations
at NLO accuracy in QCD with Shower Monte Carlo programs.  The
calculation is performed by \powhel, relying on the \powhegbox\
framework, that allows for the matching between the fixed order
computation, with input of matrix elements produced by the \helacnlo\
collection of event generators, and the Parton Shower evolution,
followed by hadronization and hadron decays as described by \pythia\
and \herwig. We focus on the dilepton and trilepton decay channels,
studied recently by the CMS Collaboration. 
}


\arxivnumber{arXiv:yymm.nnnn}

\begin{document}
\maketitle
\flushbottom

\section{Introduction}

The hadroproduction of \tT-pairs in association with vector bosons is
one of the key processes to constrain top quark properties, in
particular top couplings, and to detect if anomalies, possibly related
to physics beyond the Standard Model (SM), can manifest themselves. 
Furthermore, it can be considered a background process for new physics
searches. In particular, the dilepton decay channel with two same-sign
leptons, accompanied by missing energy and jets, is a relatively rare
channel in the SM, but largely exploited in recent supersymmetry
searches~\cite{CMSsusy}.  From the experimental point of view, these
studies are becoming feasible thanks to the increasing amount of data
collected at LHC, that has already reached an integrated luminosity
large enough to permit the disentangling of \tT~+~V signals over other
SM backgrounds~\cite{CMSnew}.  Such an investigation can certainly
benefit from high accuracy theoretical tools, involving the inclusion
of radiative corrections, at least in QCD, and the matching to Parton
Shower (PS) approaches. 

The aim of this paper is to provide predictions for \tT\ + $V$ production
(with $V$ = $W^+$, $W^-$, $Z$) at LHC at both NLO and NLO + PS accuracy. 
In case of NLO we also include uncertainties due to factorization and
renormalization scale variation, always assumed equal one to each other
in this work. This is achieved by \powhel, an event generator relying
on the \powhegbox\ \cite{Alioli:2010xd} computer framework designed 
for matching predictions at NLO accuracy in QCD to a PS evolution, 
according to the POWHEG method~\cite{Frixione:2007vw,Nason:2004rx}.
We use as input matrix elements that we compute through codes available
in the \helacnlo\ package~\cite{Bevilacqua:2011xh}.  With such an
input, the \powhegbox\ is capable to make predictions at both NLO
accuracy, and at NLO accuracy matched to a PS evolution.  We especially
concentrate on the $\sqrt{s} = 7$ and 8\,TeV energies, but the approach 
can easily be extended to other ones (and to other colliders).  By
means of this same framework we were already able to produce
predictions for other processes (\tT, \ttj, \tth/A) at the same
accuracy, which can be considered a good test of its robustness
\cite{Kardos:2011qa,Garzelli:2011vp,Garzelli:2011iu,Dittmaier:2012vm}. 
So far, we also presented some theoretical results on \ttz\ production
itself, at NLO accuracy~\cite{Kardos:2011na}, and a phenomenological
study limited to its decay channel in six jets plus missing energy, at
NLO + PS accuracy~\cite{Garzelli:2011is}.

This paper is new with respect to our previous ones, since here
for the first time we produce predictions for \ttw\ hadroproduction,
and we concentrate on the (semi)leptonic decay channels of \ttz, the
same channels that are nowadays preferred by the experimental
collaborations, as much cleaner signals can be obtained with respect to
the fully hadronic decay one.  The \ttw\ hadroproduction has already
been recently investigated by \mcfm\ at the NLO accuracy in QCD
\cite{Campbell:2012dh}.  Our study provides a completely independent
confirmation of their results at the parton level, with which we found
agreement within the quoted uncertainties. Furthermore, we give
predictions for the first time for this same process at the hadron
level, by the matching the NLO predictions to the \pythia\
\cite{Sjostrand:2006za} and \herwig\ \cite{Corcella:2002jc} Shower
Monte Carlo (SMC) programs, describing PS emissions, hadronization and
hadron decays.

The paper is organized as follows.
In Section~\ref{theory} we provide a short description of the general 
computing framework, and details on the particular issues we had 
to face for the implementation of the \ttv\ specific processes.
In Section~\ref{resultsnlo} we quote our results at NLO accuracy, and
we show the checks we did to ensure that the matching between the NLO
computation and the PS algorithm is implemented in a correct way.
In Section~\ref{phenomenology} we describe the phenomenological studies
we performed at the hadron level and we show predictions for
differential distributions both at the inclusive level and in the same
exclusive selection channels considered by the CMS Collaboration in
their data analysis. In particular, our predictions turn out to be 
compatible with the experimental data in both the trilepton and the
dilepton channels, as recorded in the recently published data analysis
at 7\,TeV.
Finally, in Section~\ref{conclusions} we draw our conclusions with
mention to future refinements of this computation. 

\section{Theoretical Framework}

\subsection{Implementation}
\label{theory}

We address the problem of matching \ttv\ ($V = Z$, $W^\pm$) production at
\nlo\ level to PS programs, to this end the POWHEG approach
\cite{Frixione:2007vw,Nason:2004rx} was chosen as implemented in
\powhegbox\ \cite{Alioli:2010xd}. While details on the implementation
of \ttz\ in this framework were recorded in our previous
papers~\cite{Kardos:2011na, Garzelli:2011is}, the following
ingredients, needed by \powhegbox,  were provided in case of \ttw\
hadroproduction:
\begin{itemize}
\item{The phase space corresponding to three massive particles in the 
final state was provided in full analogy with our previous computations
of  the \ttz\
and \tth\ processes at the same accuracy
\cite{Garzelli:2011is, Garzelli:2011vp}.}
\item
The Born and real-emission matrix elements corresponding to the
$\qq\ \qaqp\ \tq\ \taq\ \rm{W}^\pm \to 0$ and 
$\qq\ \qaqp\ \tq\ \taq\ \rm{W}^\pm\ g \to 0$
processes, respectively, with q, q$'$ $\in$ \{u, d, c, s\}, 
were provided by \helacnlo\ \cite{Bevilacqua:2011xh}.
\item
The finite part of the virtual amplitudes was computed by
\helaconeloop\ \cite{vanHameren:2009dr} for the
$\qq\ \qaqp\ \tq\ \taq\ \rm{W}^\pm \to 0$ processes.
\item
At both tree- and one-loop-level the remaining matrix elements were
obtained by crossing.
\item
The spin- and color-correlated Born squared matrix elements were also
provided by \helacnlo.
\end{itemize}
The \powhel\  (= \powhegbox\ + \helacnlo) code implemented this way is
capable of generating Les Houches Events (\lhe's), including up to
first radiation emission, for both \ttwp\ and  \ttwm. A selection
between these two cases can simply be achieved by setting the
\verb#Wmode# keyword in the input card to $\pm 1$.

In order to make comparison with the available \nlo\ predictions
\cite{Campbell:2012dh}, we had to use a non-diagonal \ckm\ matrix in
the calculation. We thus extended \helaconeloop\, in this respect. This
process can then be considered the first one, among those computed with
\helaconeloop, where a non-diagonal \ckm\ matrix was used. A check of
the correctness of the implementation was provided by comparing our
results with those already available in literature (see next
Subsection), obtained in the same non-diagonal conditions.  We make
available the \powhel\ implementation, where the user has the
possibility of switching from the diagonal  \ckm\ matrix
to a non-diagonal one by specifying a positive value of the
\verb#sin2cabibbo# keyword in the input card, which declares
$\sin^2\theta_C$. 

\subsection{Results at NLO accuracy and Checks}
\label{resultsnlo}

In order to assess the correctness of the implementation, a standard
set of checks we are used to doing on the \powhel\ implementation of
any new process, was performed also in this case.  The consistency
between the real emission matrix elements, the Born part, and the real
counterterms automatically computed according to the FKS subtraction
scheme~\cite{Frixione:1995ms}, was checked by investigating the
behavior of these terms in all kinematically degenerate regions of
phase space. The original and crossed matrix elements computed by
\powhel\ were checked against those provided by \helacphegas\ and
\helaconeloop\ standalone in various randomly chosen phase space
points. As for \ttw, the Born results were checked against \mcfm\
\cite{Campbell:2010ff,mcfmweb}, and the \nlo\ ones against the
predictions quoted in Ref.~\cite{Campbell:2012dh}, using the same set
of parameters mentioned therein and $\sin^2\theta_C = 4.9284\cdot
10^{-2}$, as in the default version of \mcfm.  In all cases we found
full agreement.  

We also compute NLO \ttw\ cross-sections at LHC for a different static
central scale choice, by considering the interval $[\mu_0/2,\, 2\mu_0]$
centered around $\mu_0 = m_t+m_V/2$, and the following set of parameters:
$\sqrt{s}$ = 7 and 8 $\tev $, the \texttt{CTEQ6.6M} \pdf\ set with a
2-loop running \as\ and 5 active flavours, taken from
\lhapdf~\cite{Whalley:2005nh}, $m_{\rm b} = 0$, whereas as for heavy
particle masses, the latest available values provided by the \pdg\
\cite{Nakamura:2010zzi},  i.e. $\mt = 173.5\,\gev$, $\mw=80.385\,\gev$
and $\mz = 91.1876\,\gev$, were adopted.  For the whole calculation a
non-diagonal \ckm-matrix was used, in the first two families, with
$\sin^2\theta_C = 4.9284\cdot 10^{-2}$. The renormalization and
factorization scales were fixed to $\mu_0$.  The predictions for the
total NLO cross-sections in these conditions are shown in
Table~\ref{tbl:nloXS}. The considered scale choice turned out to
provide a flatter scale dependence with respect to the case
$\mu_0=\mt$, as can be understood by comparing the results quoted in
Table~\ref{tbl:nloXS} to those provided in Ref.~\cite{Campbell:2012dh}. 
\begin{table}
\begin{center}
\begin{tabular}{|c|c|c|c|c|c|}
\hline
\hline
& $\sqrt{s}$ (\tev) & $\mu$ & $\sigma^\lomath$ (fb) & $\sigma^\nlomath$ (fb) & $\mathcal{K}$-fact. \bigstrut\\
\hline
\multirow{6}{*}{\ttwp} & \multirow{3}{*}{7} & $\mu_0/2$ & 121.8(1) & 114.3(1) & \multirow{3}{*}{1.13} \\
\cline{3-5}
                       &                    & $\mu_0$   & 93.1(1)  & 104.7(1) & \\
\cline{3-5}
                       &                    & $2\mu_0$  & 72.7(1)  & 93.8(1) & \\
\cline{2-6}
                       & \multirow{3}{*}{8} & $\mu_0/2$ & 159.3(1) & 156.2(2) & \multirow{3}{*}{1.16} \\
\cline{3-5}
                       &                    & $\mu_0$   & 122.9(1) & 142.6(2) & \\
\cline{3-5}
                       &                    & $2\mu_0$  & 96.7(1)  & 127.5(1) & \\
\hline
\hline
\multirow{6}{*}{\ttwm} & \multirow{3}{*}{7} & $\mu_0/2$ & 46.7(1) & 46.9(1) & \multirow{3}{*}{1.20} \\
\cline{3-5}
                       &                    & $\mu_0$   & 35.6(1) & 42.6(1) & \\
\cline{3-5}
                       &                    & $2\mu_0$  & 27.8(1) & 38.0(1) & \\
\cline{2-6}
                       & \multirow{3}{*}{8} & $\mu_0/2$ & 64.1(1) & 67.1(1) & \multirow{3}{*}{1.23} \\
\cline{3-5}
                       &                    & $\mu_0$   & 49.4(1) & 60.5(1) & \\
\cline{3-5}
                       &                    & $2\mu_0$  & 38.9(1) & 53.9(1) & \\
\hline
\hline
\multirow{6}{*}{\ttz}  & \multirow{3}{*}{7} & $\mu_0/2$ & 141.6(1) & 149.4(2) & \multirow{3}{*}{1.32} \\
\cline{3-5}
                       &                    & $\mu_0$   & 103.5(1) & 136.9(1) & \\
\cline{3-5}
                       &                    & $2\mu_0$  & 77.8(1)  & 120.8(1) & \\
\cline{2-6}
                       & \multirow{3}{*}{8} & $\mu_0/2$ & 209.5(1) & 224.9(4) & \multirow{3}{*}{1.34} \\
\cline{3-5}
                       &                    & $\mu_0$   & 153.9(1) & 205.7(2) & \\
\cline{3-5}
                       &                    & $2\mu_0$  & 116.2(1) & 181.7(2) & \\
\hline
\hline
\end{tabular}
\caption{\label{tbl:nloXS} 
\powhel\ predictions for the inclusive \ttwp, \ttwm\ and \ttz\
cross-sections at \lo\ and \nlo\ QCD accuracy at LHC for $\sqrt{s}$ = 7
and 8\,TeV, for various static scale choices, centered around $\mu_0$ =
$m_t$ + $m_V/2$, with V = W for the \ttw\ cases and $Z$ for the \ttz\
one. The statistical uncertainties of our simulations are shown 
in parentheses.}
\end{center}
\end{table}

\begin{figure}[t]
\begin{center}
\includegraphics[width=.49\textwidth]{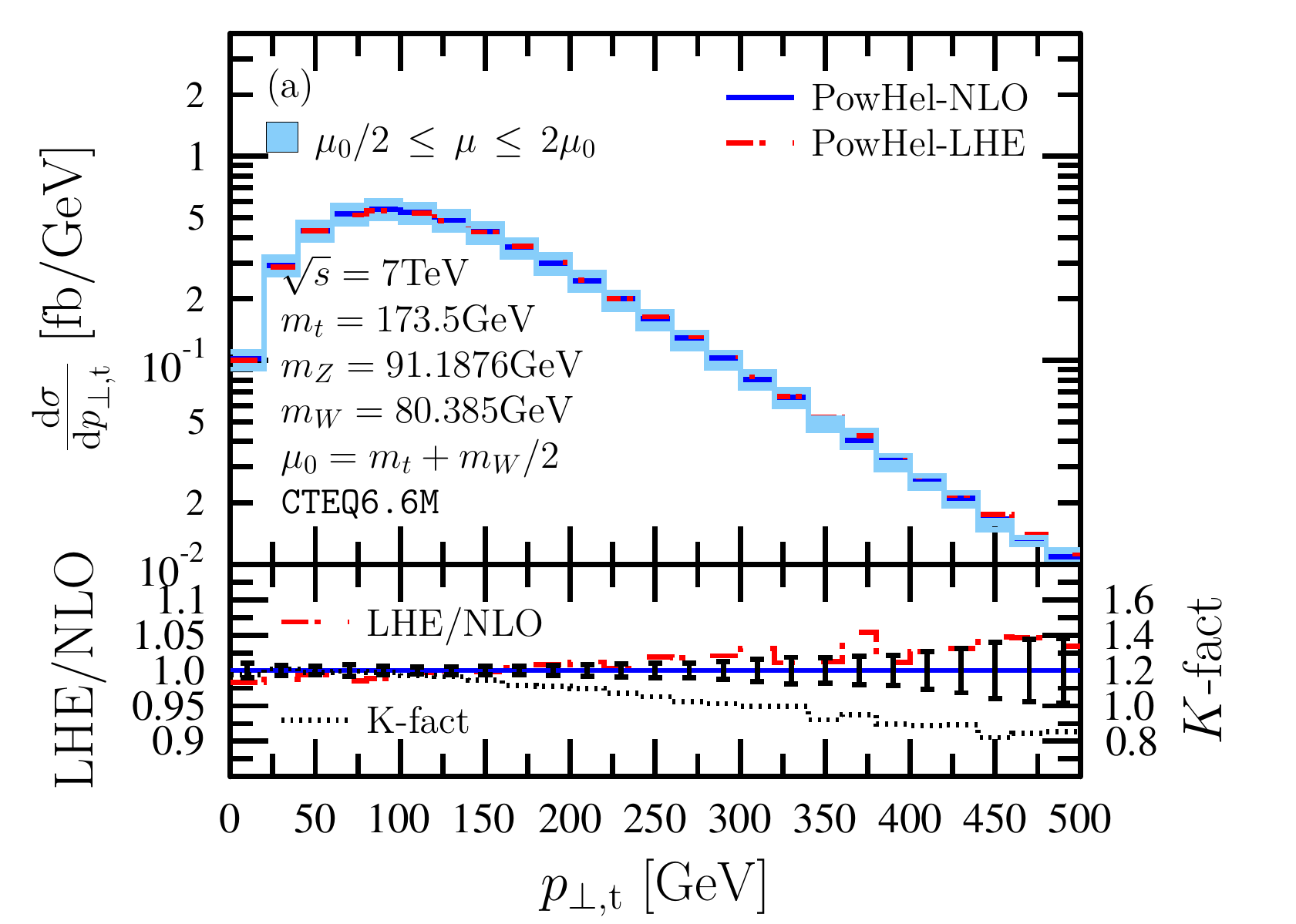} 
\includegraphics[width=.49\textwidth]{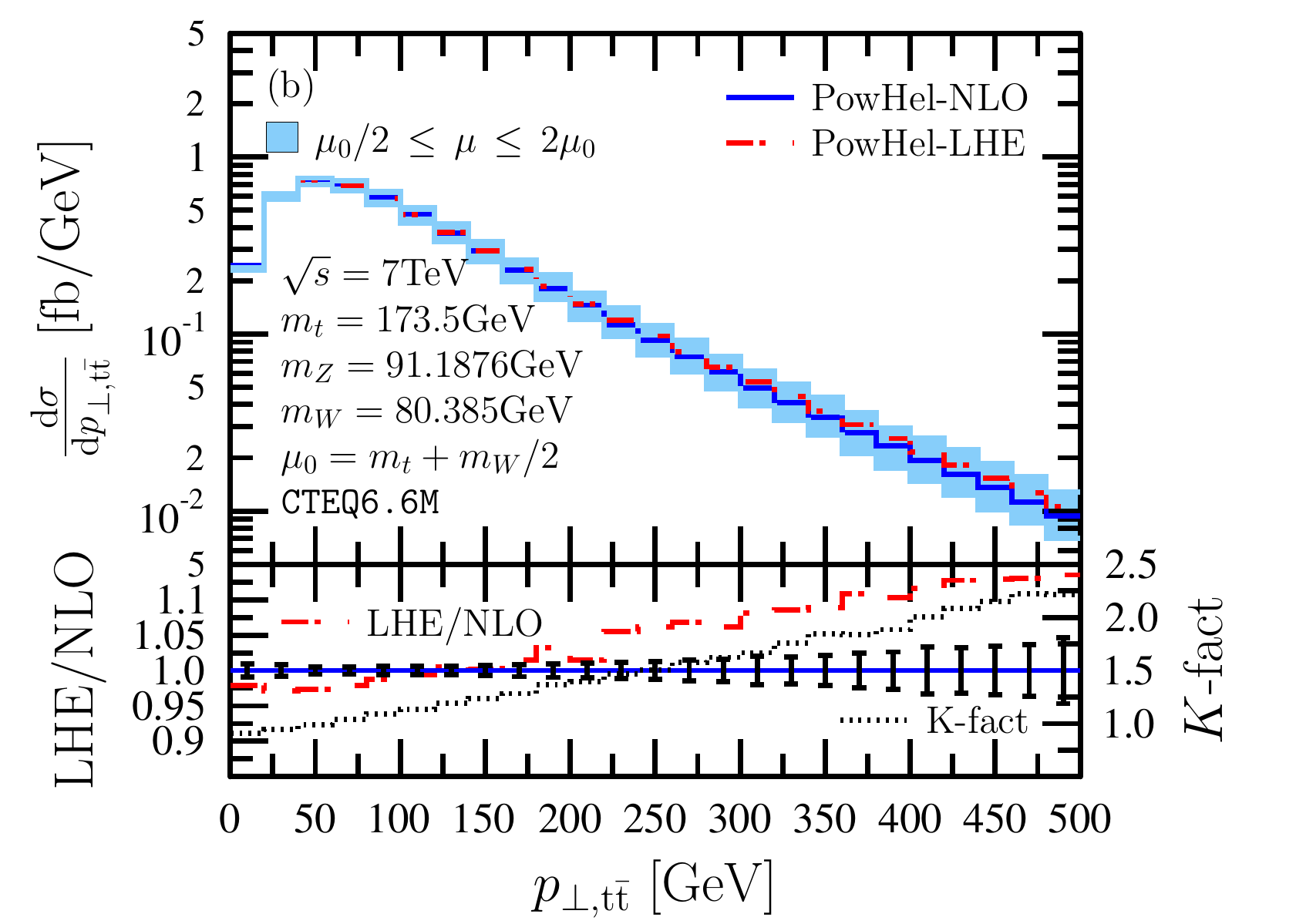}
\includegraphics[width=.49\textwidth]{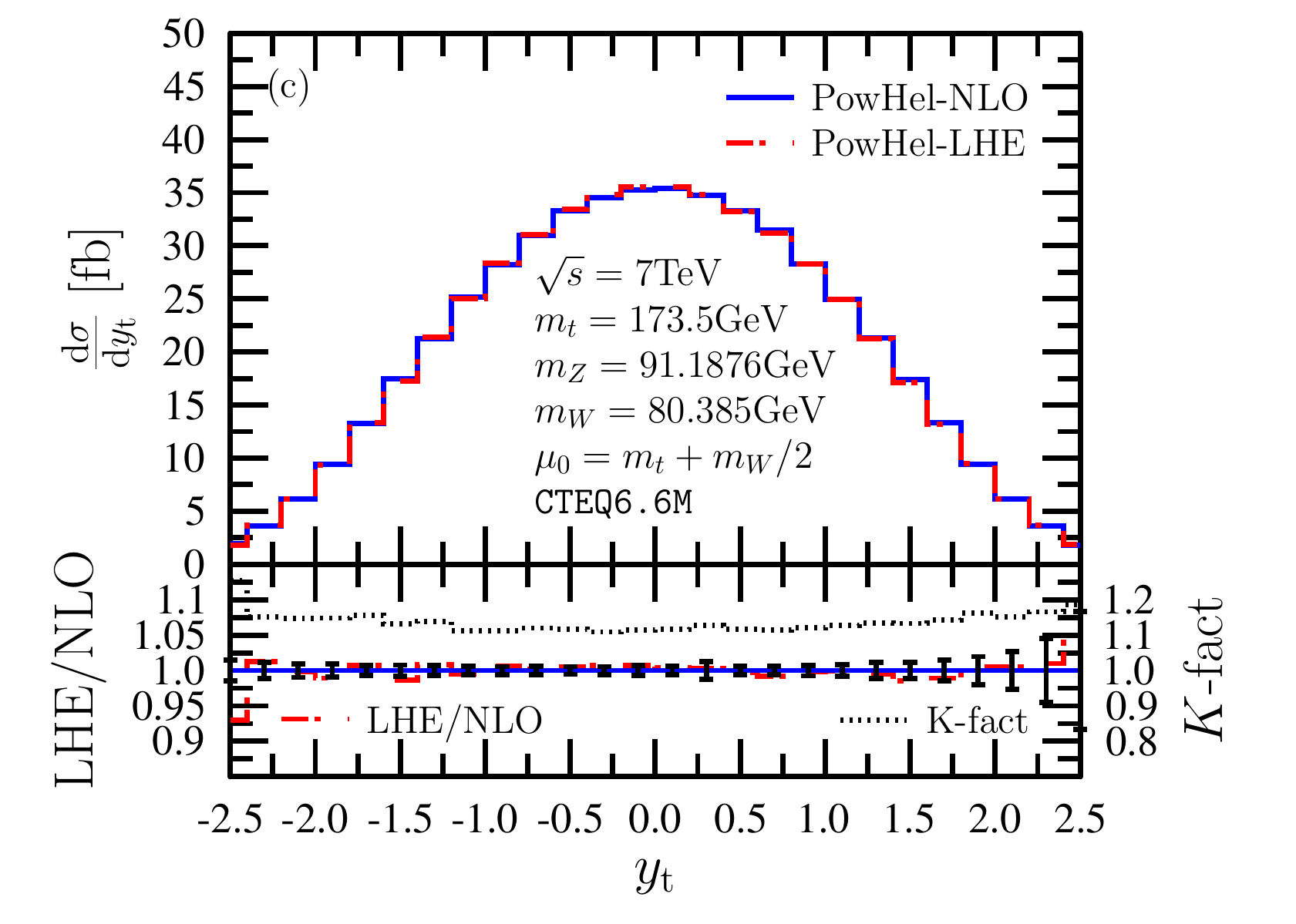} 
\includegraphics[width=.49\textwidth]{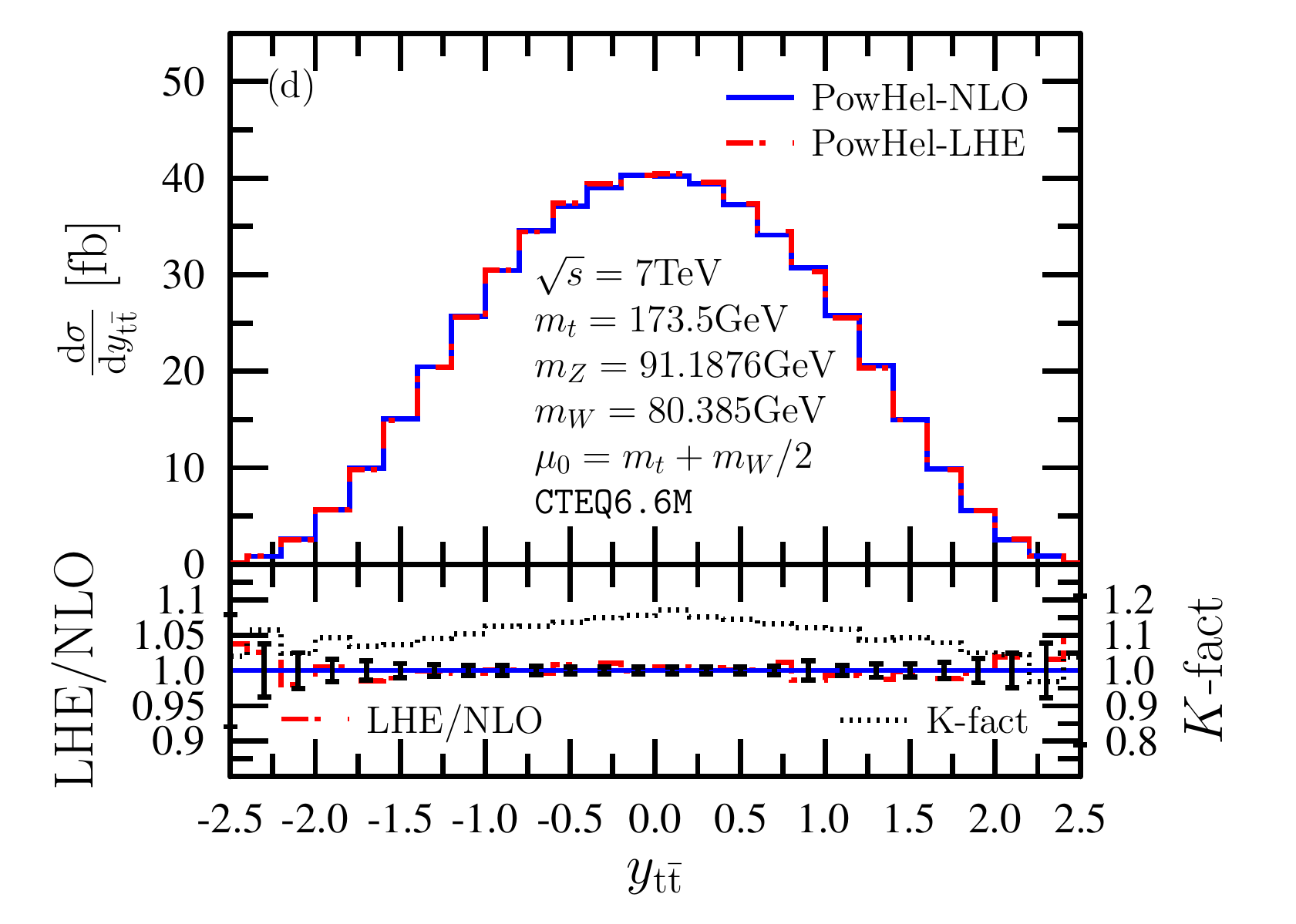}
\end{center}
\caption{\label{fig:nlocomp_ttwp}
Comparison between differential distributions at \nlo\ accuracy (solid
line) and from the LHE's (dashed line), in case of \ttwp\ production.
As sample distributions, the transverse momentum and rapidity
distributions are shown for the t-quark and for the \tT-pair. 
In the lower panels the red dash-dotted line corresponds to the LHE/\nlo\
ratio, whereas the differential K-factor (NLO/LO) is depicted 
with a dotted line. The error-bars refer to the statistical
uncertainties on the LHE/\nlo\ ratio. In case of transverse momentum
distributions, the scale dependence is also superimposed as a
light-blue band, which represents a scale variation between $\mu_0/2$
and $2\mu_0$.}
\end{figure}

Although the K-factor associated to the \ttw\ process is close to one,
it is also informative to compare \nlo\ differential cross-sections to
those obtained from the LHE's, which checks the correctness of the
matching procedure. Sample distributions can be found in
\figs{fig:nlocomp_ttwp}{fig:nlocomp_ttwm}, where the transverse momenta
and the rapidities of both the t-quark and the \tT-pair are shown in
case of \ttwp\ and \ttwm, respectively, together with the ratio of the
predictions from the LHE's to the NLO ones. The agreement between the
\nlo\ and the LHE distributions is quite remarkable, as can be seen
from the two rapidity plots and from the \pt distribution of the
t-quark. The small deviation visible in the \ptt\ tail is within the
increased statistical uncertainty in that region, also plotted in the
lower inset of each panel. For the \ptttbar\ distribution the agreement
is within 5\,\% up to $\simeq 220$\,GeV, but worsens in the high momentum
tail. We attribute this increasing difference to the increasing K-factor
that reaches 2 around 400\,GeV (also depicted in the lower panel of the
plot). This 10\,\% deviation however, is well within the NLO scale
dependence, as seen from the upper panel, where the uncertainty band,
corresponding to a scale-variation in the $[\mu_0/2, 2 \mu_0]$ interval,
is shown as well.

\begin{figure}[t]
\begin{center}
\includegraphics[width=.49\textwidth]{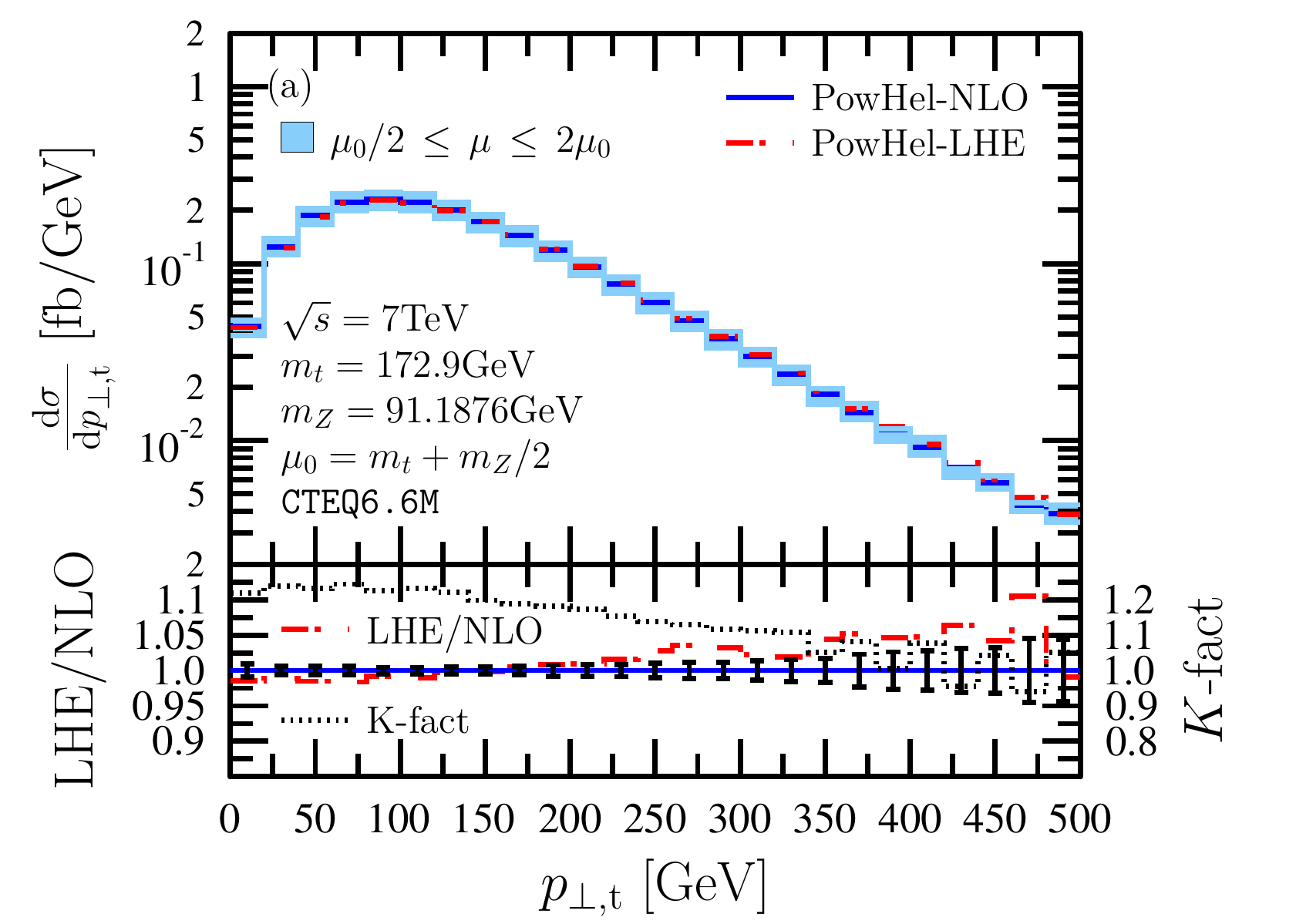} 
\includegraphics[width=.49\textwidth]{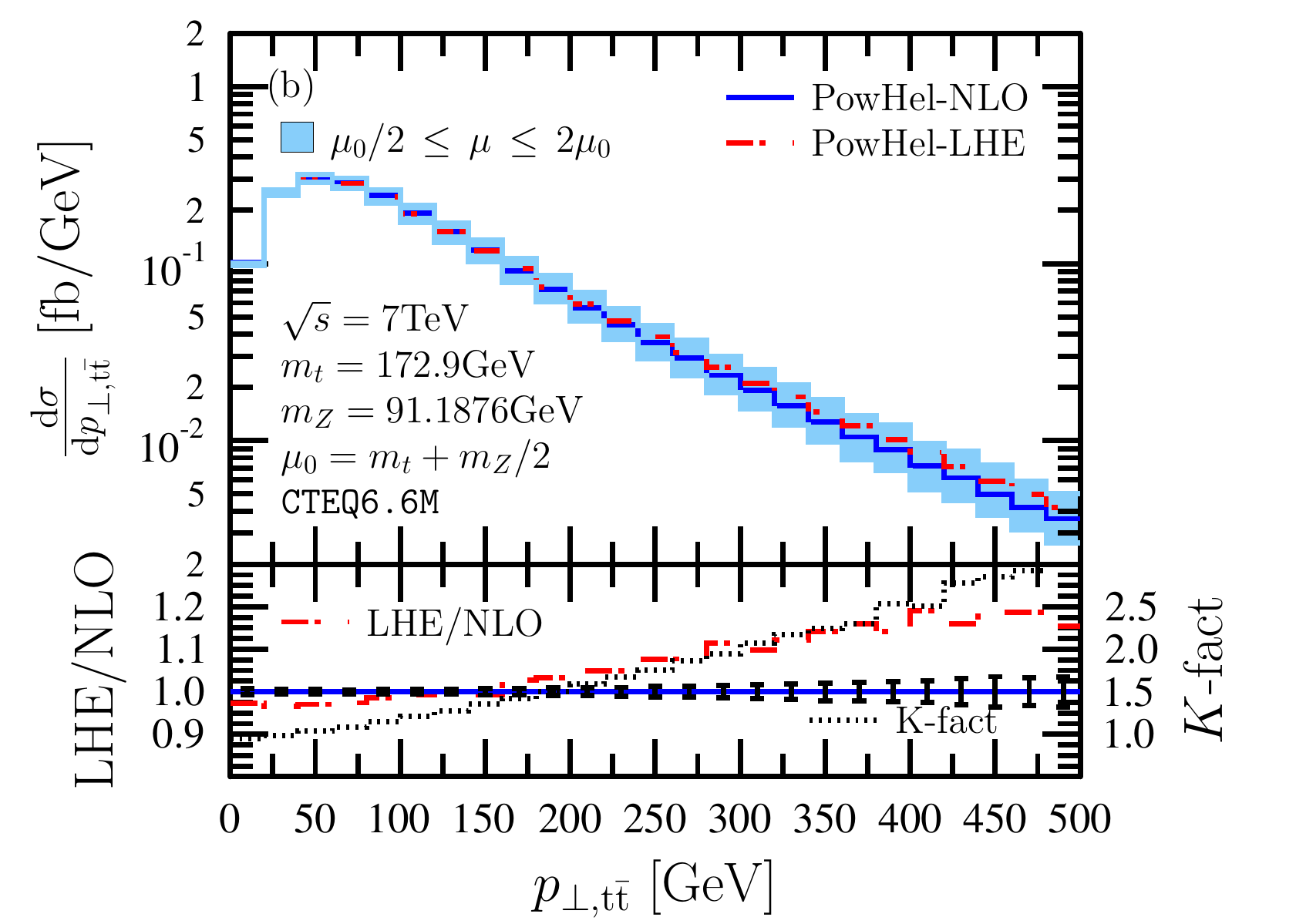}
\includegraphics[width=.49\textwidth]{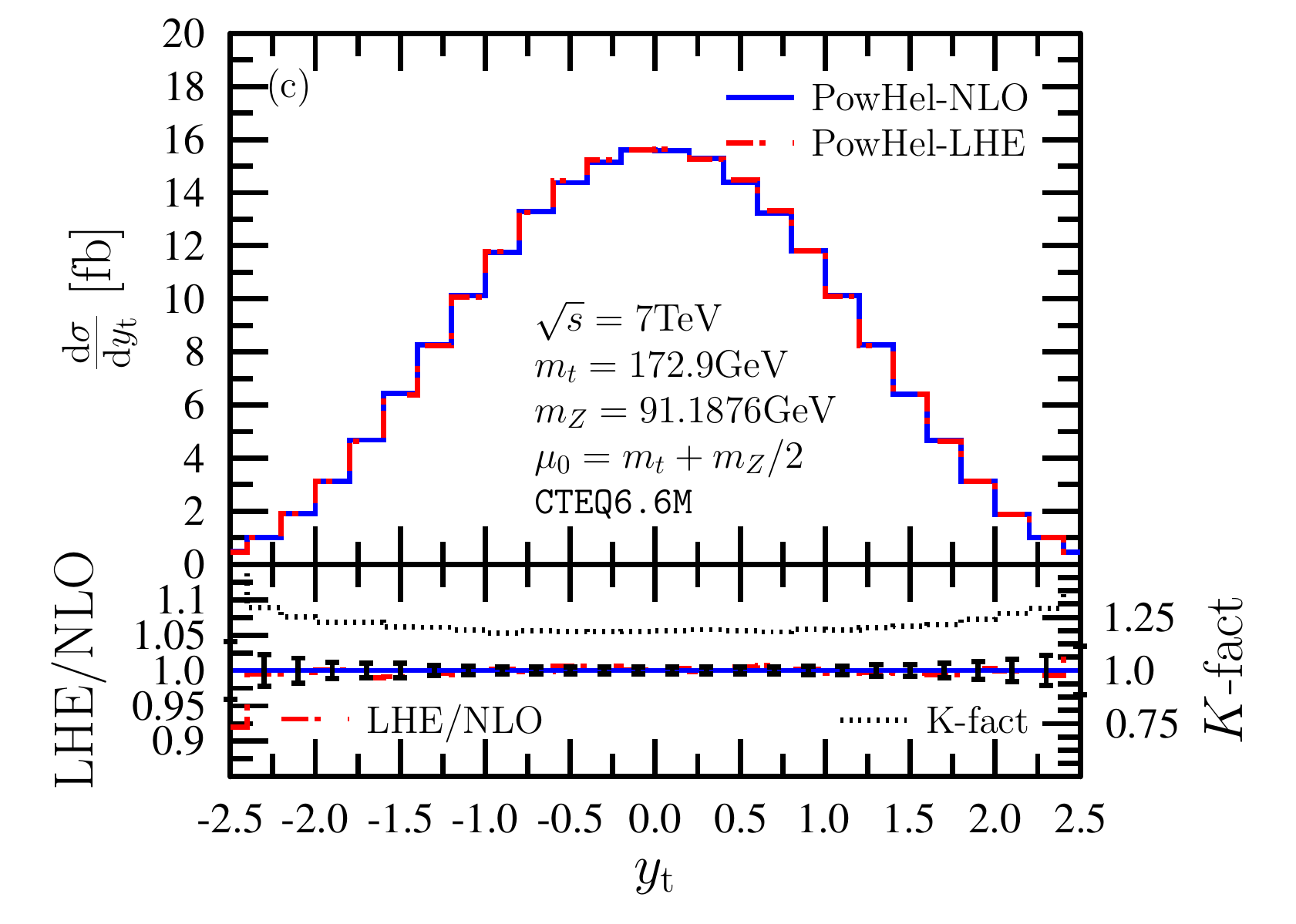} 
\includegraphics[width=.49\textwidth]{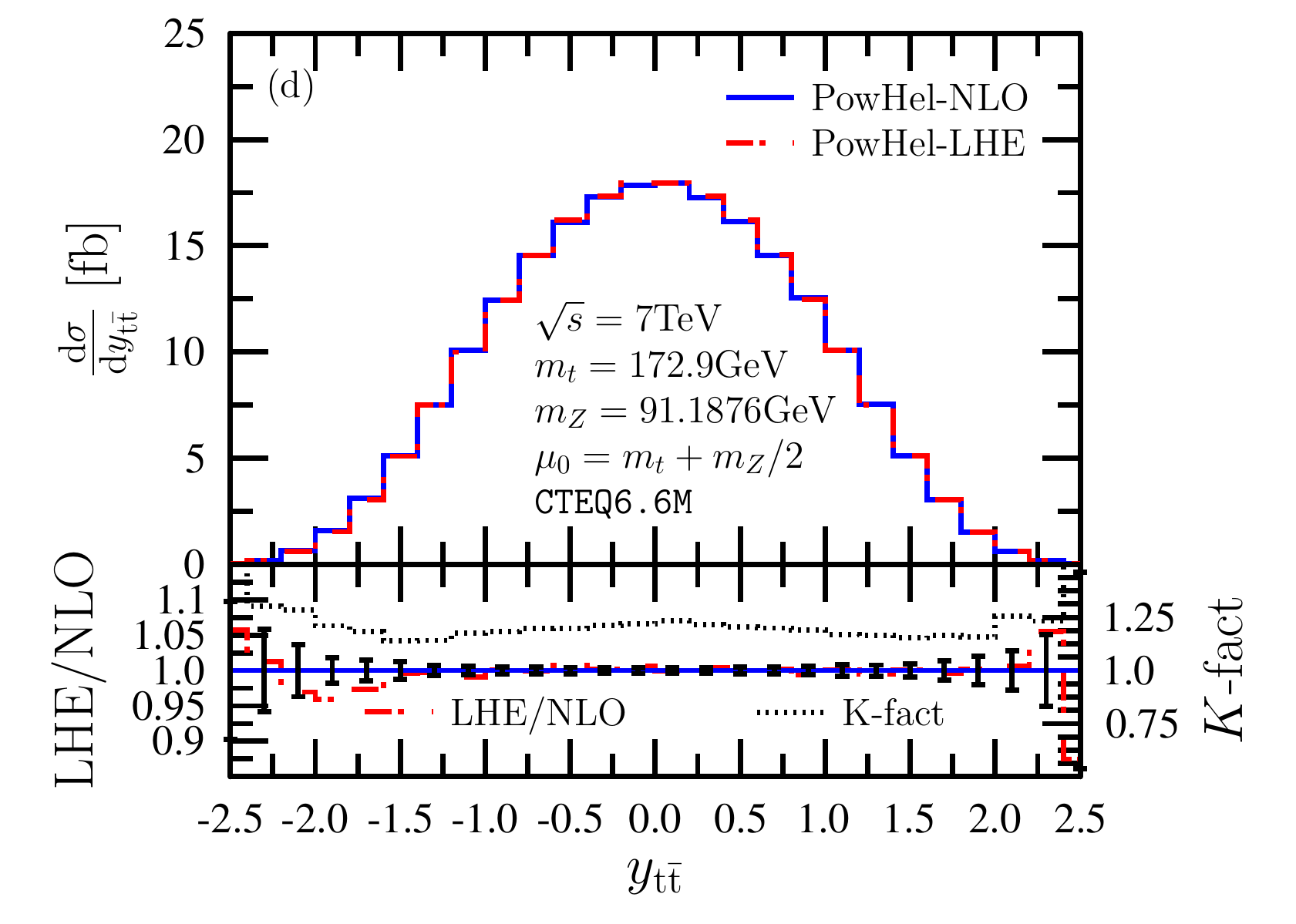} 
\end{center}
\caption{\label{fig:nlocomp_ttwm} Same as Fig.~\ref{fig:nlocomp_ttwp}, 
as for \ttwm\ production.}
\end{figure}

Differential K-factors and the comparison between NLO and LHE
distributions in case of the \ttz\ process can be found in
Ref.~\cite{Kardos:2011na, Garzelli:2011is} 

\section{Phenomenology}
\label{phenomenology}

\subsection{PowHel and SMC setup}
\label{setup}

For our phenomenological studies the following parameters were adopted
in \powhel: the \texttt{CTEQ6.6M} \pdf\ set, with a 2-loop running \as,
$\mt=172.5\,\gev$, $\mw=80.385\,\gev$, $\mz = 91.1876\,\gev$, 
$\sin^2\theta_C = 4.9284\cdot 10^{-2}$.  The renormalization and
factorization scales were fixed to $\mur=\muf=\mt + m_V/2$.  Although
the value of $m_{\rm t}$ is different from the most recent measurements
at the LHC and also from that used in our NLO comparisons, it was also
used in Ref.~\cite{Campbell:2012dh} and in several measurements performed
by the LHC experiments so far.

The \powhel\ code provides collection of LHE's of two kinds: Born-like
events, and events including first radiation emission. According to the
POWHEG method, this emission is SMC independent.  Further emissions can
be simulated by simply showering the events by SMC programs, under the
condition that the first emission remains the hardest. We consider the
last fortran version of both the \pythia\ and \herwig\ SMC, providing a
virtuality-ordered and an angular-ordered PS, respectively. 
As the ordering variable in the POWHEG method is the relative transverse
momentum, in case of an angular-ordered PS parton emissions with larger
transverse momentum than the first one have to be vetoed explicitly (done
in \herwig\ automatically). Furthermore, a truncated shower, simulating
wide-angle soft emission before the hardest one ought to be included, too.
However, the effect of the truncated shower in general turns out to be
small, as shown e.g. in Ref.~\cite{LatundeDada:2006gx} and as we
already verified in case of many different multiparticle production
processes including a \tT\ pair, where the predictions of \pythia\ and
\herwig\ turn out to agree one with each other within a few percent. 
Thus, we neglect truncated shower contributions in this paper, as we
already did in our previous ones.  

These SMC codes were also used to generate t-quark and heavy boson
decays (neglecting spin correlations), as well as hadronization and
hadron decays.  For consistency, heavy particle masses in the SMC setup
were set to the same values used in the \powhel\ computation, whereas
the light quark masses in \herwig\ were set to the default values
implemented in \pythia.  Heavy particle decay widths were fixed to
$\Gamma_t = 1.45775$\,GeV, $\Gamma_W = 2.085$\,GeV\ and $\Gamma_Z  =
2.4952$\,GeV.  Decays of heavy bosons into electrons were assumend to
have the same branching ratio as into muons.  $\pi^0$'s were enforced
to be stable in both SMC's, as they can be easily reconstructed in the
experiments from their decay products (2 $\gamma$), and muon stability
was enforced in \herwig, as in \pythia\ default configuration. All
other particles and hadrons were assumed to be stable or to decay
according to the default implementation of each SMC.
Multiple interactions were neglected in both SMC's.

\subsection{Inclusive analysis}
\label{inclu}

We now present predictions at the SMC level, i.e. after PS,
hadronization and hadron decay, in case of \ttwp, \ttwm\ and \ttz\, in
the most general case, i.e.  without applying any selection cut. This
is possible since these processes are finite at the Born level, so we
did not have to introduce any technical cut in the \powhel\ generation
of LHE's.  It is useful and instructive to present some theoretical
distributions at this level, to better understand how the selection
cuts that we will discuss in the following will modify these predictions.   
In particular, we focus on a few selected distributions that will also 
be shown again,  in presence of cuts, in the following Subsections.  

The inclusive cross-sections at the SMC level are the same as at the
NLO level as the POWHEG method ensures that the cross-sections from
LHE's coincide with the exact NLO ones, i.e.
$\sigma_{\rm {LHE}} = \sigma_{\rm {NLO}}$. We found that
$\sigma_{\ttz}  > \sigma_{\ttwp} > \sigma_{\ttwm}$, 
with $\sigma_{\ttz}$ = 137.21 $\pm$  0.01~fb,
$\sigma_{\ttwp}$ = 106.74 $\pm$ 0.01~fb and
$\sigma_{\ttwm}$~=~43.472~$\pm$~0.005~fb, respectively (uncertainties are
statistical only).  These values are slightly larger than those quoted
in Table~\ref{tbl:nloXS}, due to the slightly smaller value of the
t-quark mass (see the beginning of the previous Subsection~\ref{setup}). 

\begin{figure}
\includegraphics[width=.49\textwidth]{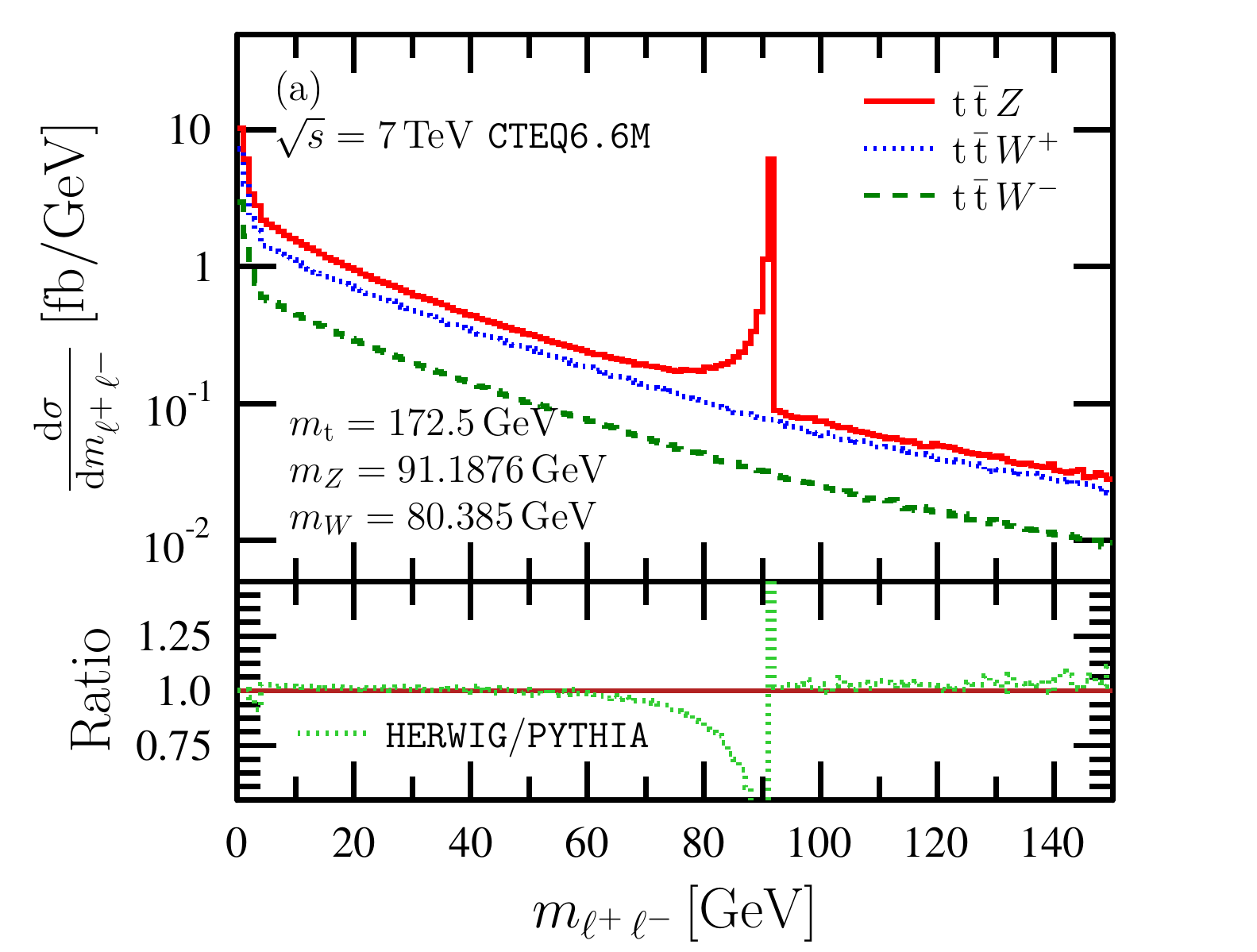} 
\includegraphics[width=.49\textwidth]{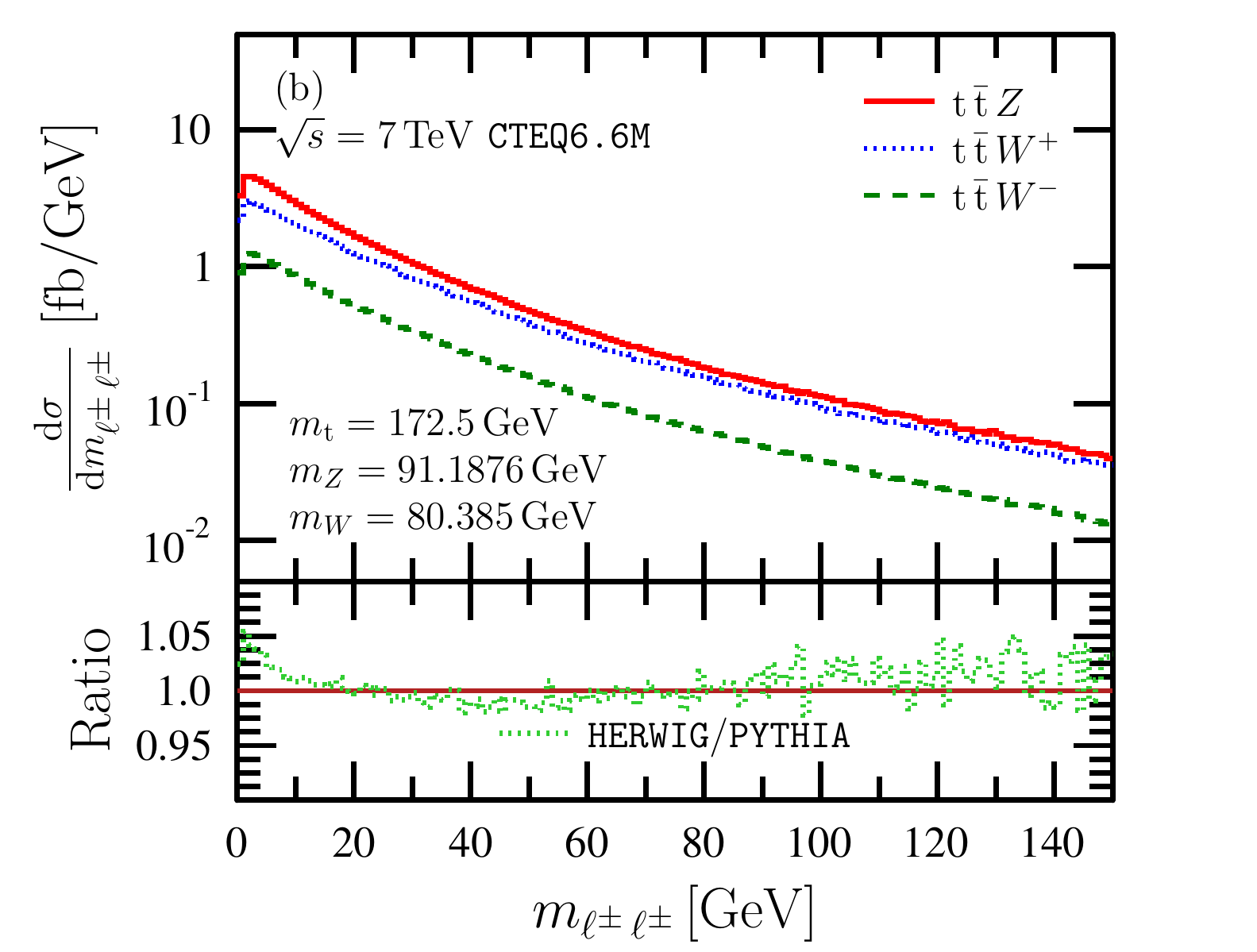} 
\caption{\label{invmassnocut} Invariant mass of
a) all ($\ell^+$, $\ell^-$) same-flavour lepton-antilepton pairs and
b) all ($\ell$,~$\ell$) same-sign lepton and anti-lepton pairs from all
events in the inclusive analysis, as obtained by \powhel~+~\pythia\ at
the $\sqrt{s} = 7$\,TeV LHC.  Predictions for the three processes \ttz,
\ttwp, and \ttwm\  are shown separately.
In the lower panel, the ratio between the cumulative predictions 
of \powhel\ + \herwig\ and \powhel\ + \pythia\ is also shown.}
\end{figure}

The invariant mass of all same-flavour ($\ell^+$, $\ell^-$) pairs in
all events is plotted in Fig.~\ref{invmassnocut}.a. Even in absence of
cuts, a peak is well visible in the \ttz\ distribution, around the $Z$
pole mass, due to $Z \to \ell^+ \ell^-$ decays.  The $e^+ e^-$ and
$\mu^+ \mu^-$ channels both contribute with a similar shape to this
distribution. The presence of this peak, absent in the \ttwp\ and
\ttwm\ distributions also plotted in Fig.~\ref{invmassnocut}.a, will be
exploited in the trilepton analysis discussed in the following
\sect{trilepton}. Looking at the invariant mass of all
same-flavour same-sign (anti-)lepton pairs in all events, plotted in
Fig.~\ref{invmassnocut}.b, an almost monotonically decreasing
distribution is found.  These lepton combinations can come from any
possible source: one from the (anti-)t-quark and the other from the $W$
or $Z$, a prompt and a secondary (anti-)leptons, two secondary
(anti-)leptons. 

The predictions using \herwig\ as SMC, instead of \pythia, agree
with the \pythia\ ones well below 5\,\% in all the dilepton mass range
considered (see the ratios plotted in both lower panels of
Fig.~\ref{invmassnocut}), except for a small window in the range
between 65 and 92\,GeV in case of opposite-sign and same-flavour
dilepton pairs, where a big difference is found, due to the different
physics implemented in the default version of the two codes: while
\pythia\ includes photon bremsstrahlung from leptons, this effect is
absent in \herwig. Thus, the sharp peak seen by \herwig\ due to
Z $\rightarrow$ ($\ell^+$, $\ell^-$) decays, is smeared in case of
\pythia. We will return on this point in more details in the following
\sect{trilepton}.  
\begin{figure}
\includegraphics[width=.49\textwidth]{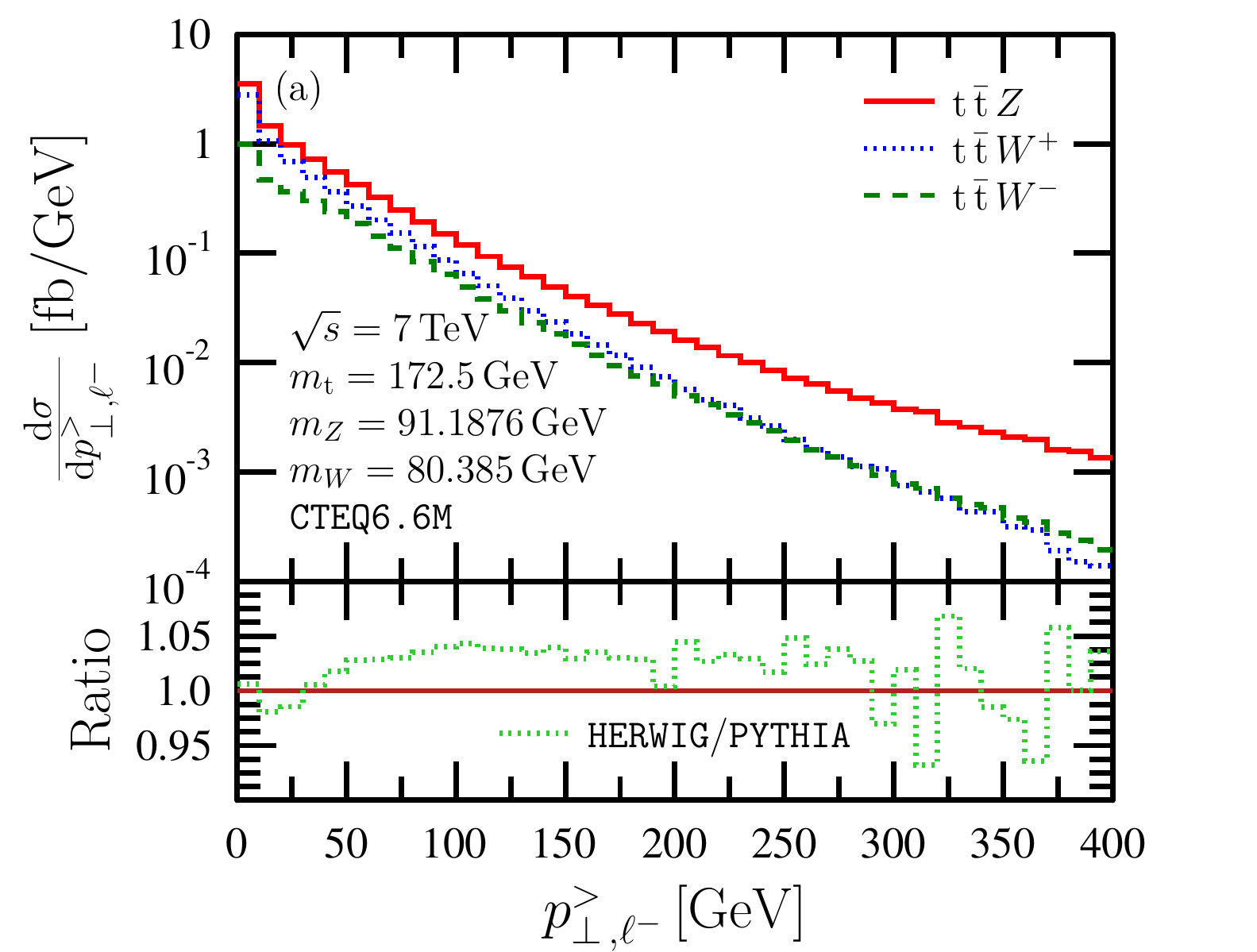}
\includegraphics[width=.49\textwidth]{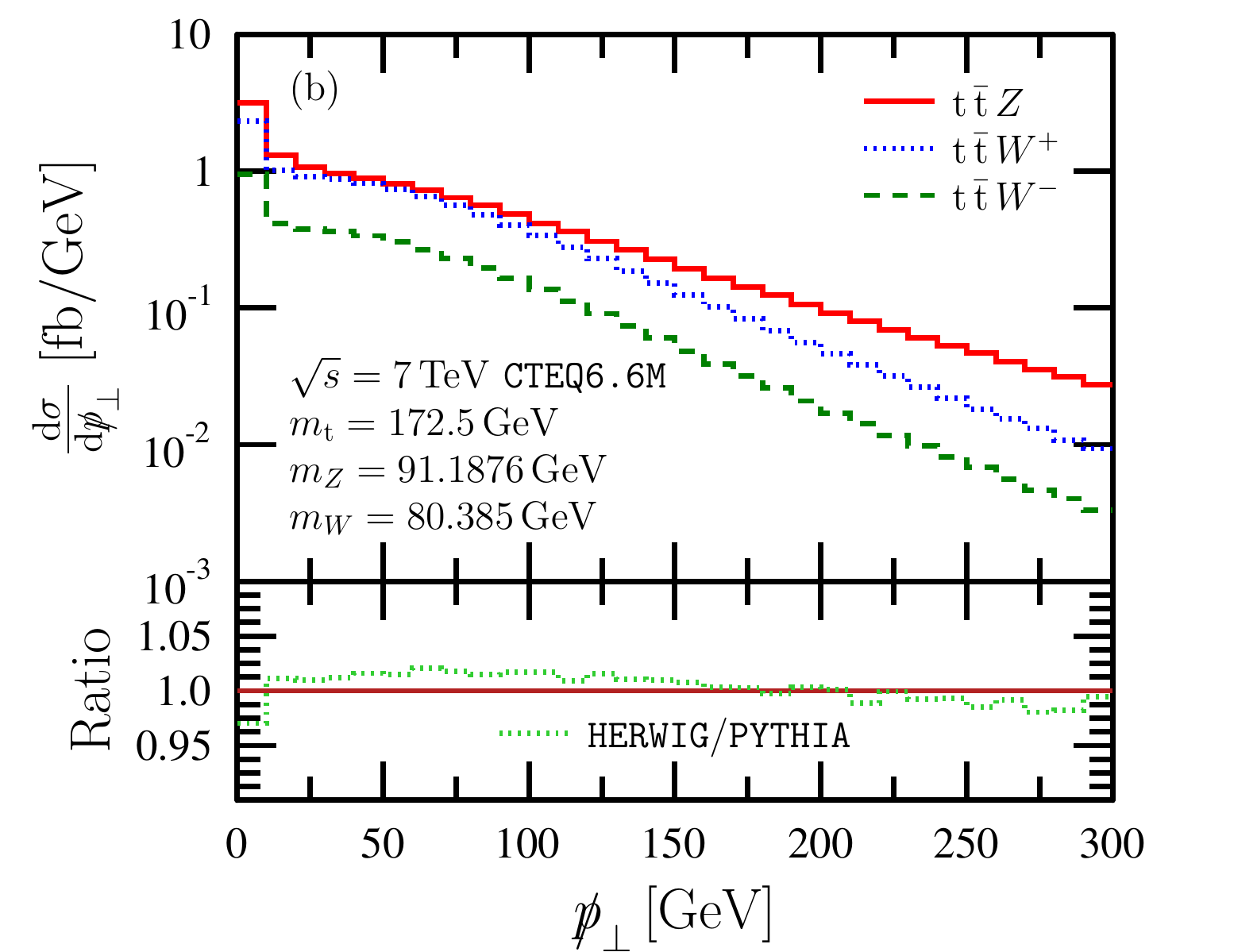} 
\caption{\label{ptmissnocut} Distributions of
a) the transverse momentum of the hardest lepton and
b) the missing transverse momentum due to all neutrinos from all
events in the no-cut analysis, as obtained by \powhel\ + \pythia\ at
the $\sqrt{s} = 7$\,TeV LHC.  Predictions for the three processes 
\ttz, \ttwp, and \ttwm\ are shown separately.
In the lower panel, the ratio between the cumulative predictions of 
\powhel\ + \herwig\ and \powhel\ + \pythia\ is also shown.}
\end{figure}

In Fig.~\ref{ptmissnocut}.a, the transverse momentum distribution of
the hardest lepton of each event is shown. Here it is worth noting the
different shapes of the \ttwp\ and \ttwm\ distributions, with the
\ttwm\ becoming larger than the \ttwp\ one for $p_\bot > 260$\,GeV, as
expected because the high $p_\bot$ tail is populated by prompt leptons
emitted from  primary $W^- \to \ell \, \nu_\ell$ decays, that are
absent in case of $W^+$ decays. Leptons originated by primary $Z$
decays can have even larger $p_\bot$ as seen from the shape of the tail
of the \ttz\ distribution, with a slope flatter than both the previous ones.  

Finally, the missing transverse momentum distribution due to all
neutrinos is plotted in Fig.~\ref{ptmissnocut}.b.  The shape of the
\ttwp\ distribution is similar to the \ttwm\ one, with a rescaling
factor just due to the different cross-section, whereas the shape of
the \ttz\ distributions differs from the previous ones, with a larger
contribution in the first two bins, due to events without neutrinos or
with neutrinos from secondary decays with very small transverse energy
and a flatter slope than the \ttw\ cases. The region around 50\,GeV,
where the the \ttwp\ and \ttz\ distributions are closer together,
is filled by neutrinos from prompt $W^+$ decays, absent in case of
\ttz. The first bin is enhanced in all distributions due to the
possibility of events without neutrinos ($W$ decays in two light jets are
indeed possible and not ruled out by any selection cut in this analysis). 

For both distributions plotted in Fig.~\ref{ptmissnocut} we found
that the differences between the cumulative predictions by \pythia\ and
\herwig, obtained by summing over the three \ttv\ processes, are within
5\,\% (see the lower panels), with a slightly larger agreement in case
of the \pTmiss-distribution.

\subsection{Trilepton-channel analysis}
\label{trilepton}

The aim of the trilepton channel analysis proposed in Ref.~\cite{CMSnew}
is selecting \ttz\ events, with Z decaying in two opposite-sign charged
leptons, and one of the quarks of the \tT-pair decaying leptonically,
whereas the other one hadronically. In particular, 
we considered the following set of cuts:
\begin{enumerate}
\item
at least two opposite-charge, same-flavor leptons with 
$p_{\bot,\,\ell} > 20$\,GeV and within CMS acceptance ($|\eta_\ell| <$
2.4, with an additional cut on the electrons impinging on the
barrel/endcap transition region of the electromagnetic calorimeter
(ECAL), corresponding to the pseudorapidity interval 1.4442 $<
|\eta_\ell| <$ 1.566),
\item
constrain the invariant mass of the dilepton system (``reconstructed
$Z$'') within the 81\,GeV/$c^2 < m_{\ell\ell} < 101$\,GeV/$c^2$ interval,
\item
$p_{\bot,\,{\ell\ell}}> 35$\,GeV, where $p_{\bot,\,{\ell\ell}}$ is the 
transverse momentum of the reconstructed $Z$,
\item
at least a third lepton in the event with $p_{\bot,\,\ell_3} > 10$\,GeV
and obeying the same pseudorapidity requirements as the other two leptons,
\item
at least three jets with $p_{\bot,\,j} > 20$\,GeV\ and $|\eta_j| < 2.4$,
of which two positively b-tagged,
\item
$H_{\rm T} >  120$\,GeV, defined as the  scalar sum of the transverse
momenta of all jets with $p_{\bot,\,j} > 20$\,GeV and  $|\eta_j| < 2.4$.
\end{enumerate}

In our simulation, jets were reconstructed using the anti-$k_\bot$
algorithm, with $R=0.5$, using \fastjet\  \cite{Cacciari:2011ma}.
b-tagging was done by means of the \texttt{MCTRUTH} parameter, allowing
to trace back the origin of a jet to a b or a $\bar{\rm b}$ quark. 
In case of multiple dilepton pairs with opposite charge and same
flavour satisfying cuts 1), 2) and 3), the pair with the invariant mass
closest to the nominal Z mass was selected. 

\begin{wrapfigure}{r}{0.6\textwidth}
\includegraphics[width=0.6\textwidth]{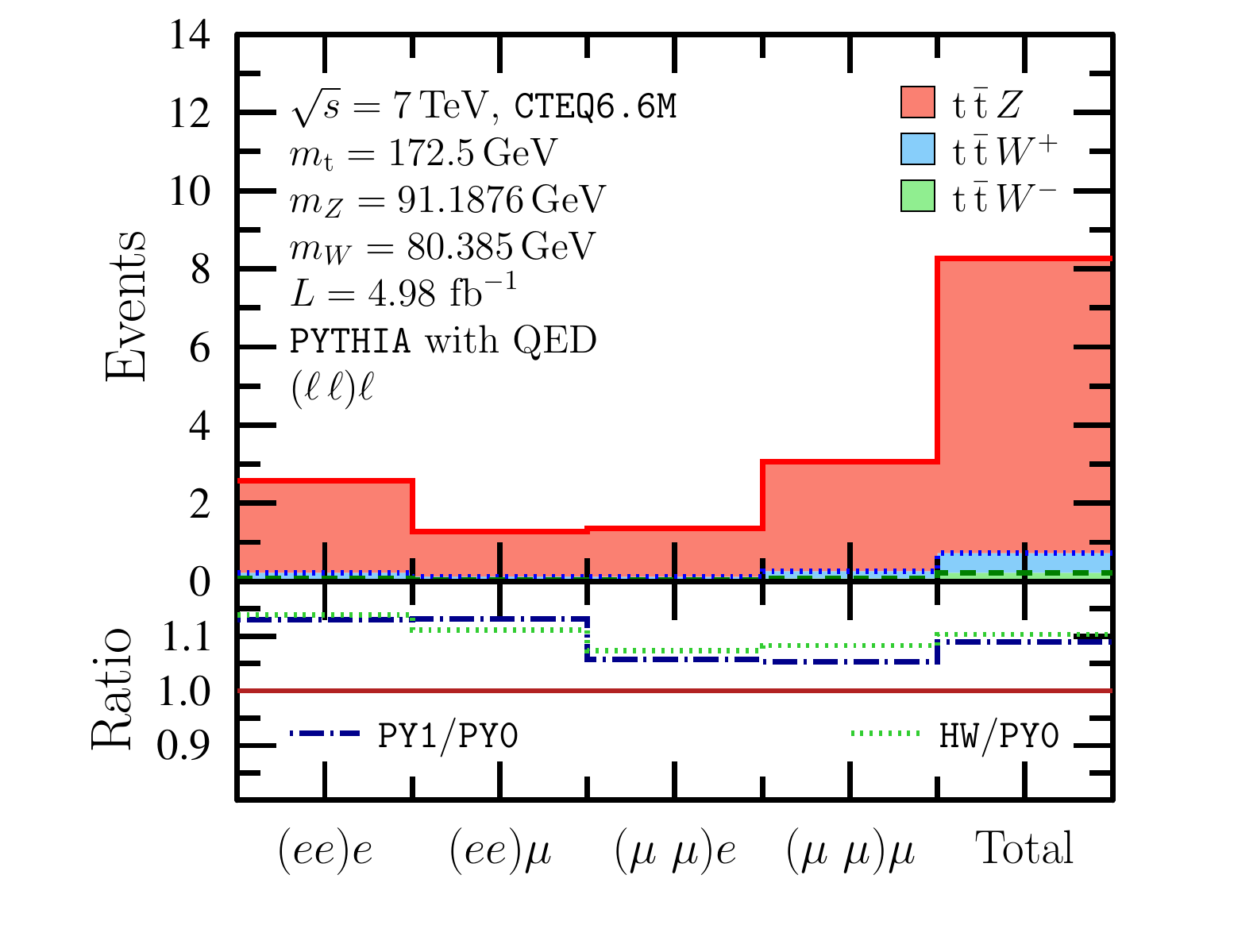}
\caption{\label{figeventstri}
Number of events in the trilepton channels at the $\sqrt{s} = 7$\,TeV LHC,
as predicted by \powhel\ +~\pythia, for an integrated luminosity
amounting to $L = 4.98$\,fb$^{-1}$. The contribution in the ($e$, $e$)
$e$,  ($e$, $e$) $\mu$, ($\mu$, $\mu$) $e$ and ($\mu$, $\mu$) $\mu$
channels are shown separately, as well as their sum in the last bin.  
The contributions due to \ttz, \ttwp\ and \ttwm\ are cumulated one over
the other. To be compared with the experimental data in Fig.~4 of the
CMS technical report~\cite{CMSnew}.  In the lower inset the ratios
between cumulative results using different SMC (\herwig/\pythia) and
between cumulative results obtained by neglecting and including photon
bremsstrahlung from leptons (\pythia-no-brem/\pythia) are also shown. 
}
\end{wrapfigure}

Predictions for the expected number of events after cuts at the
$\sqrt{s} = $ 7 TeV LHC, corresponding to an integrated luminosity 
$L = 4.98$\,fb$^{-1}$, as obtained by our \powhel + \pythia\ simulations,
are shown in Fig.~\ref{figeventstri}, distinguishing the possible decay
channels, labelled by the flavours of the two leptons entering the
dilepton system plus the third lepton mentioned in cut 4). When more
than one additional lepton satisfies cut 4), we choose that with the
largest $p_\bot$. The sum of the results in all channels is plotted in
the last bin of the figure, as well. These predictions can be compared
to the experimental results, presented in Ref.~\cite{CMSnew} for the
same luminosity, with the caveat that we still do not include the
predictions for background processes (like $Z$ + jets, \tT\ and diboson
production) at the same accuracy. For an estimate of these background
contributions at a lower accuracy, one can rely on Ref.~\cite{CMSnew}.
One has also to take into account that the CMS Collaboration used an
experimental b-jet tagging algorithm, instead of a purely theoretical
one, as we did.  
 
\begin{figure}[h!]
\includegraphics[width=.49\textwidth]{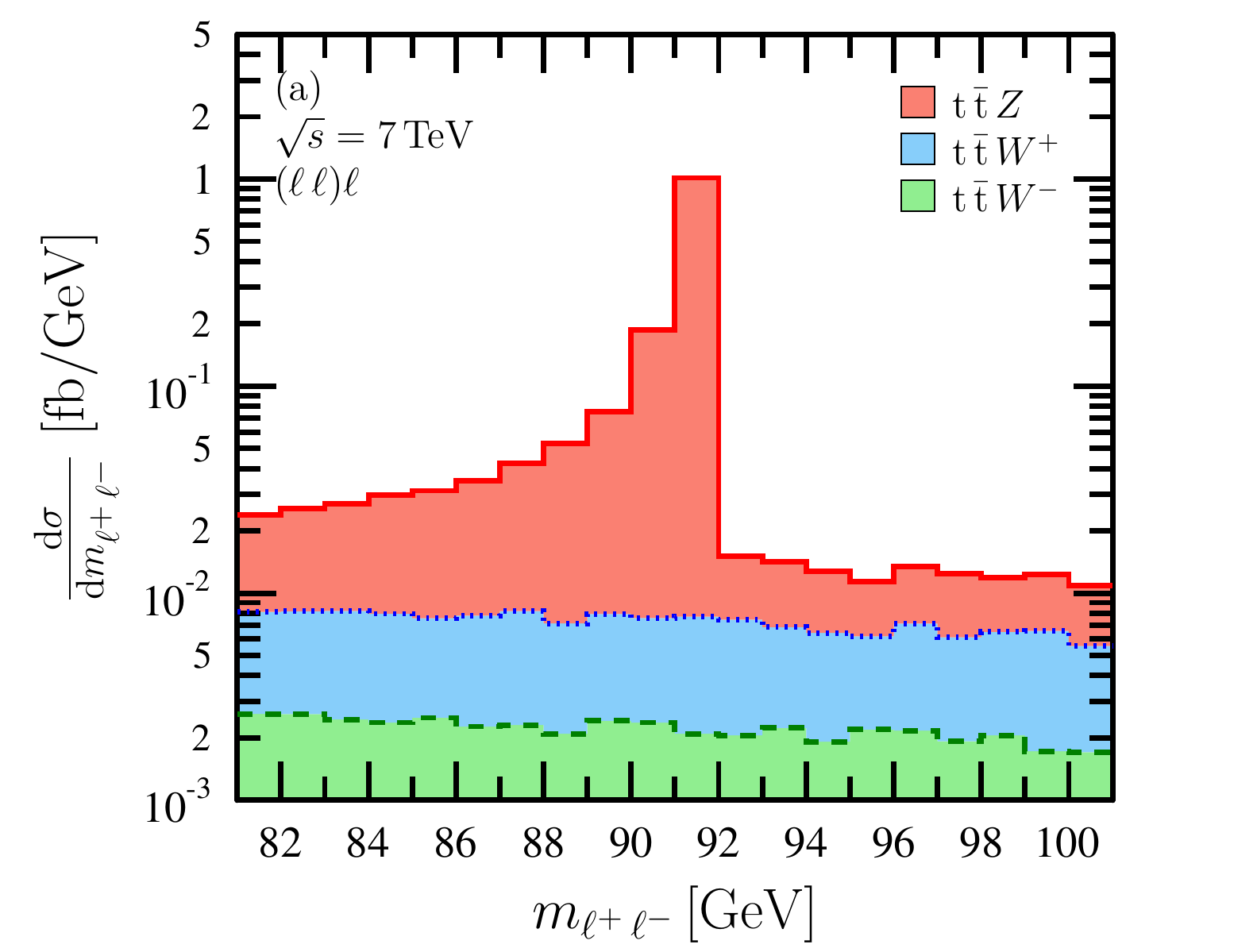}
\includegraphics[width=.49\textwidth]{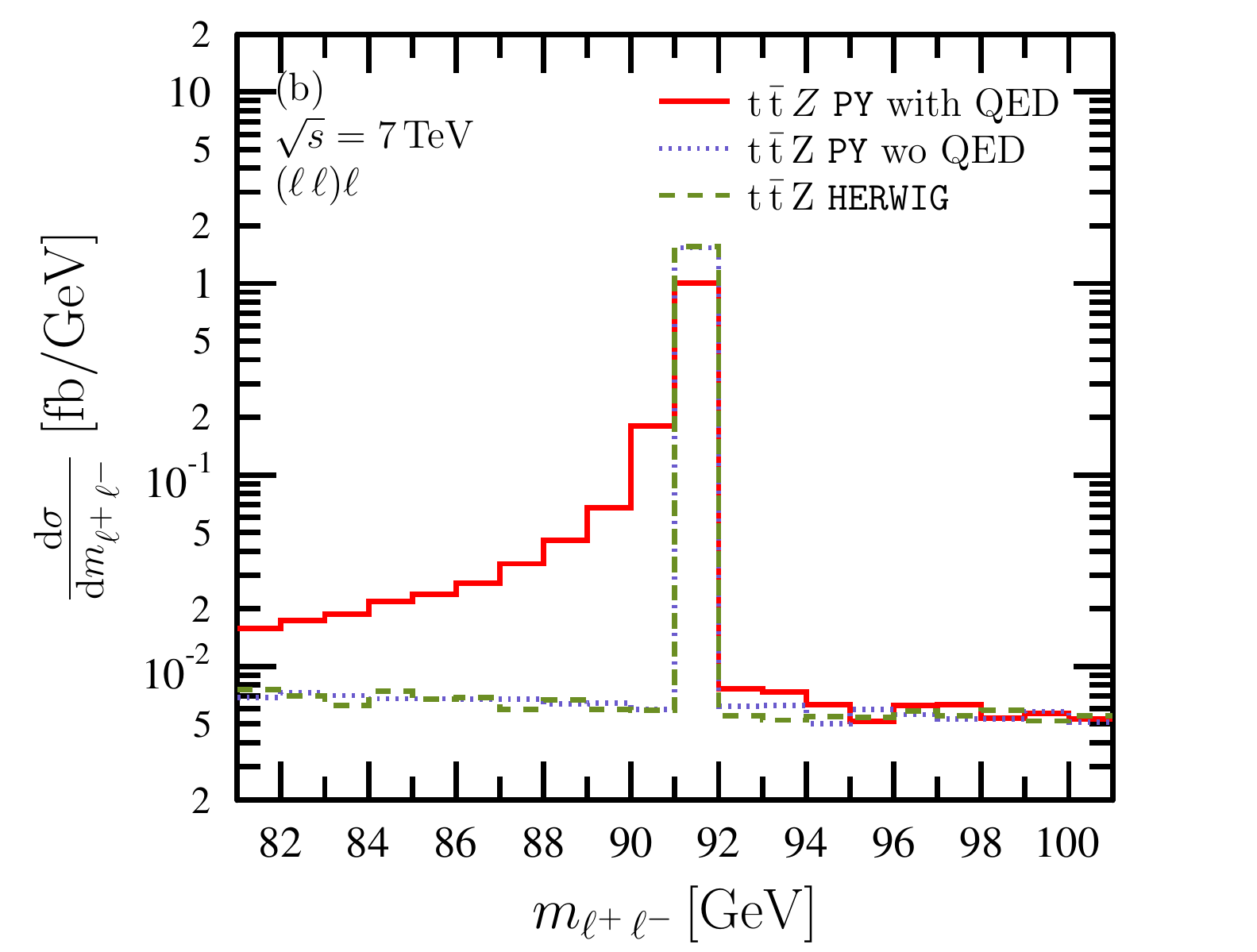}
\caption{\label{dilinvmass} Invariant mass of the $Z$ reconstructed from
same-flavour ($\ell^+$, $\ell^-$) pairs after the trilepton analysis,
as obtained by \powhel + \pythia\ at the $\sqrt{s} = 7$\,TeV LHC.
a) Predictions corresponding to the different processes  \ttz, \ttwp\
and \ttwm\ cumulated one over the other,
b) distributions obtained by using different SMC (\pythia, \herwig\ and
\pythia\ without photon bremmstrahlung from leptons) are also shown,
limited to \ttz-production.}
\end{figure}

As expected, as a result of the selection cuts, and in particular of
the cut on the invariant mass of the dilepton system, both in the
experiment and in our theoretical predictions the contributions to the
total number of events due to the \ttw\ processes are highly
suppressed. We estimate a suppression factor of about 10 between the
cross-sections after the cuts for the processes (\ttwp + \ttwm) and \ttz,
from our theoretical simulations.  The invariant mass of the
reconstructed $Z$ is plotted for these three processes in
Fig.~\ref{dilinvmass}.a, from where it is clear that the largest
contribution of the \ttz\ process is due to the peak around $m_Z$,
completely absent in case of both \ttwp\ and \ttwm.  

In the lower inset of Fig.~\ref{figeventstri} the ratios of the results
using different SMC's are plotted.  In particular, using \herwig\
instead of \pythia\ as SMC, leads to a larger number of events. This is
due to the different physics implemented in the two SMC's. In fact, as
already mentioned in \sect{inclu}, while \pythia\ includes by default
photon bremsstrahlung from leptons, the stand-alone fortran version of
\herwig\ does not include it (unless one interfaces it with external
packages). The photon bremsstrahlung effect affects the dilepton
invariant mass after SMC: as shown in Fig.~\ref{dilinvmass}.b, (that is
the analogous of Fig.~\ref{invmassnocut}.a after cuts), the very narrow
peak evident in case of \herwig\ simulations is smeared by the effect
of photon bremsstrahlung from leptons implemented in default \pythia\
simulations. As a further check, we switched off this kind of emissions
even in \pythia. The predictions of \pythia\ without lepton
bremsstrahlung are superimposed on the same plot and look to be closer
to the \herwig\ ones. The modification on the number of events after
cuts in the different channels, one gets by switching off this effect
in \pythia, is also shown in the lower panel of Fig.~\ref{figeventstri}.

The predictions presented in Fig.~\ref{figeventstri} are compatible
with the experimental data of Ref.~\cite{CMSnew} within the error-bars 
of the latter. However, while our simulations predict almost symmetric
central values between the ($e$, $e$) $e$ and the ($\mu$, $\mu$) $\mu$
channels, and between the ($e$, $e$) $\mu$ and the ($\mu$, $\mu$) $e$
channels, the experimental data show the same pattern for the latter
case, but different for the first one: the ($\mu$, $\mu$) $\mu$ bin is
more populated than the ($e$, $e$) $e$ one (although the populations of
these two bins can still can still be viewed equal within the large
error-bars). From our simulations we verified that a slight asymmetry
between the ($e$, $e$) $e$ and the ($\mu$, $\mu$) $\mu$ bins is
generated by the inclusion of photon bremsstrahlung from leptons. In
the absence of this effect, the population of these two channels
is instead completely symmetric. It is also affected by the different
selection cuts on electros and muons (see cut 1).  We think that the
larger asymmetry effects, as inferred from the experimental data, are
due to other experimental details, like limited detection efficiencies
and charge misidentification effects. Such effects are neglected in our
simulations and their precise implementation is dependent on the
experimental analysis detail, beyond the scope of this work.  

Our predictions for the cross-section contributions in the different 
trilepton channels (see Fig.~\ref{figeventstri}), summing over the three
processes \ttz, \ttwp\  and \ttwm, in case of $\sqrt{s} = 7$\,TeV LHC, 
are as follows:
$\sigma_{(e,e),e} = 0.516$\,fb,
$\sigma_{(e,e),\mu} = 0.255$\,fb,      
$\sigma_{(\mu,\mu),e} = 0.273$\,fb,
$\sigma_{(\mu,\mu),\mu} = 0.613$\,fb,
$\sigma_{{\scriptscriptstyle \sum}} = 1.658$\,fb,
all with a statistical uncertainty below 10$^{-5}$\,fb.

\begin{figure}[h!]
\includegraphics[width=.49\textwidth]{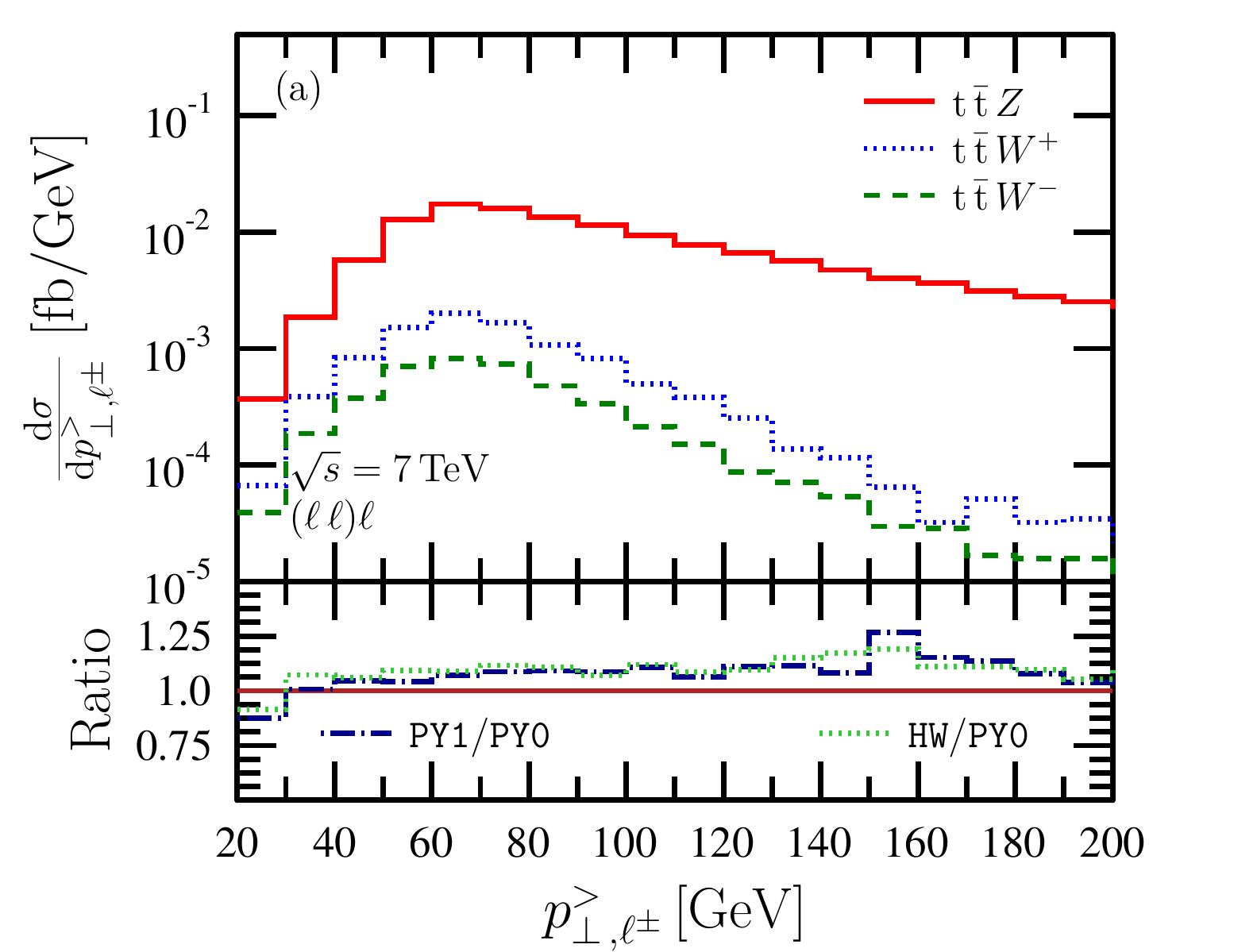}
\includegraphics[width=.49\textwidth]{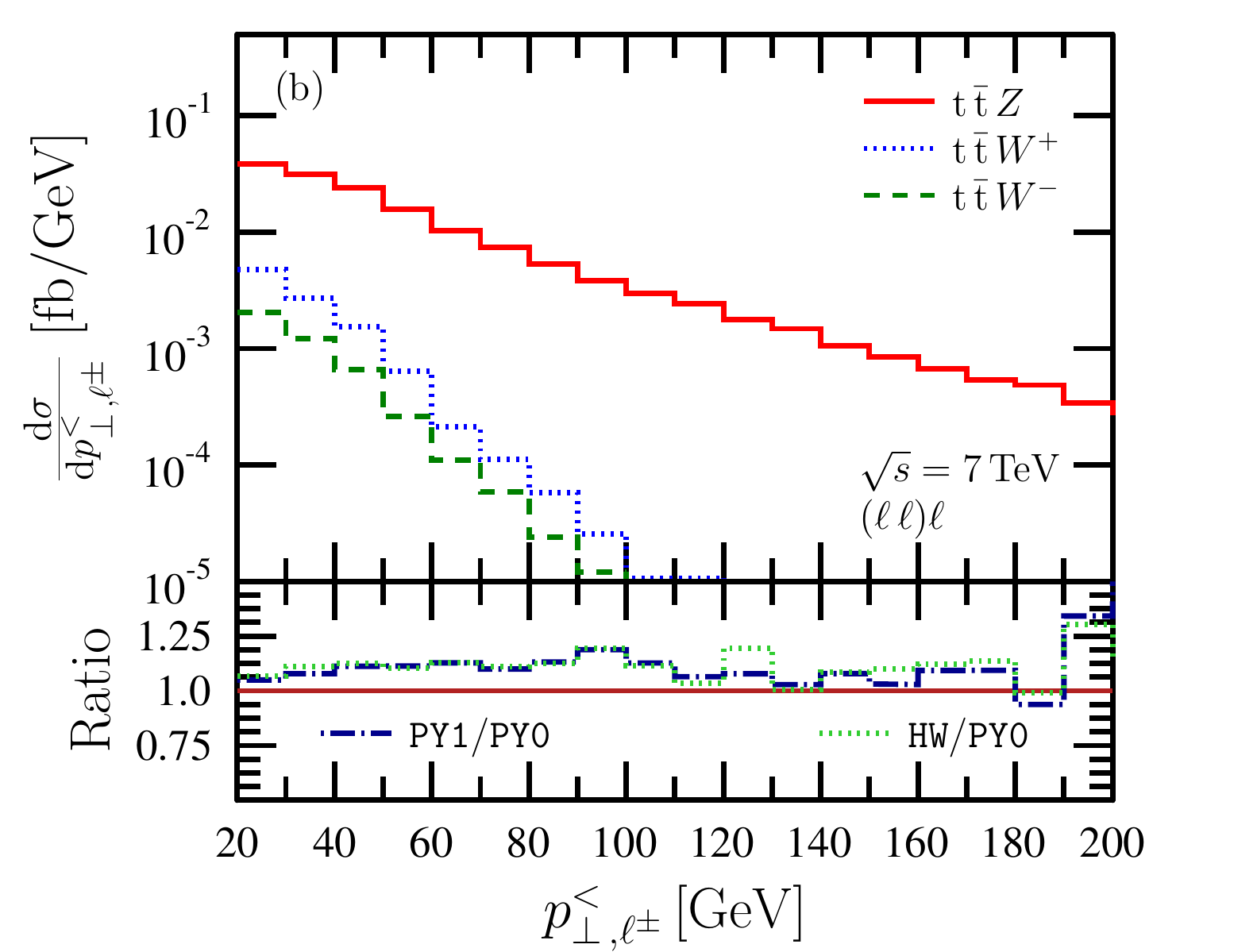}
\caption{\label{ptleadsub} Transverse momentum distributions of
a) the leading and
b) the subleading (anti-)lepton 
of each ($\ell^+$, $\ell^-$) pair corresponding to a reconstructed $Z$ boson.
Predictions by \powhel~+~\pythia, corresponding to the different \ttz,
\ttwp and \ttwm\ processes are shown separately.
In the lower inset the ratios between cumulative results using
different SMC \herwig\ and \pythia\ (\hw/\py0) and between cumulative
results obtained by neglecting and including photon bremsstrahlung from
leptons in \pythia\ (\py1/\py0) are also shown. 
}
\end{figure}

The transverse momentum distributions of the leading and subleading
(anti-)lepton of the $(\ell^+,\ell^-)$ pairs selected by the considered
system of cuts are shown separately in Fig.~\ref{ptleadsub}.  These
distributions have different shapes, as expected: those belonging to
the leading lepton are peaked at $\sim$ 65 GeV for both \ttz, \ttwp\
and \ttwm\, while those belonging to the subleding lepton decrease
monotonically just above the $p_{\bot,l} > 20$\,GeV cut.  When
considered together, the lepton and the anti-lepton give rise to a
``reconstructed $Z$'', whose $p_\bot$  has a shape characterized by a
smooth peak in the 50\,GeV region. 

\begin{figure}[h!]
\includegraphics[width=0.49\textwidth]{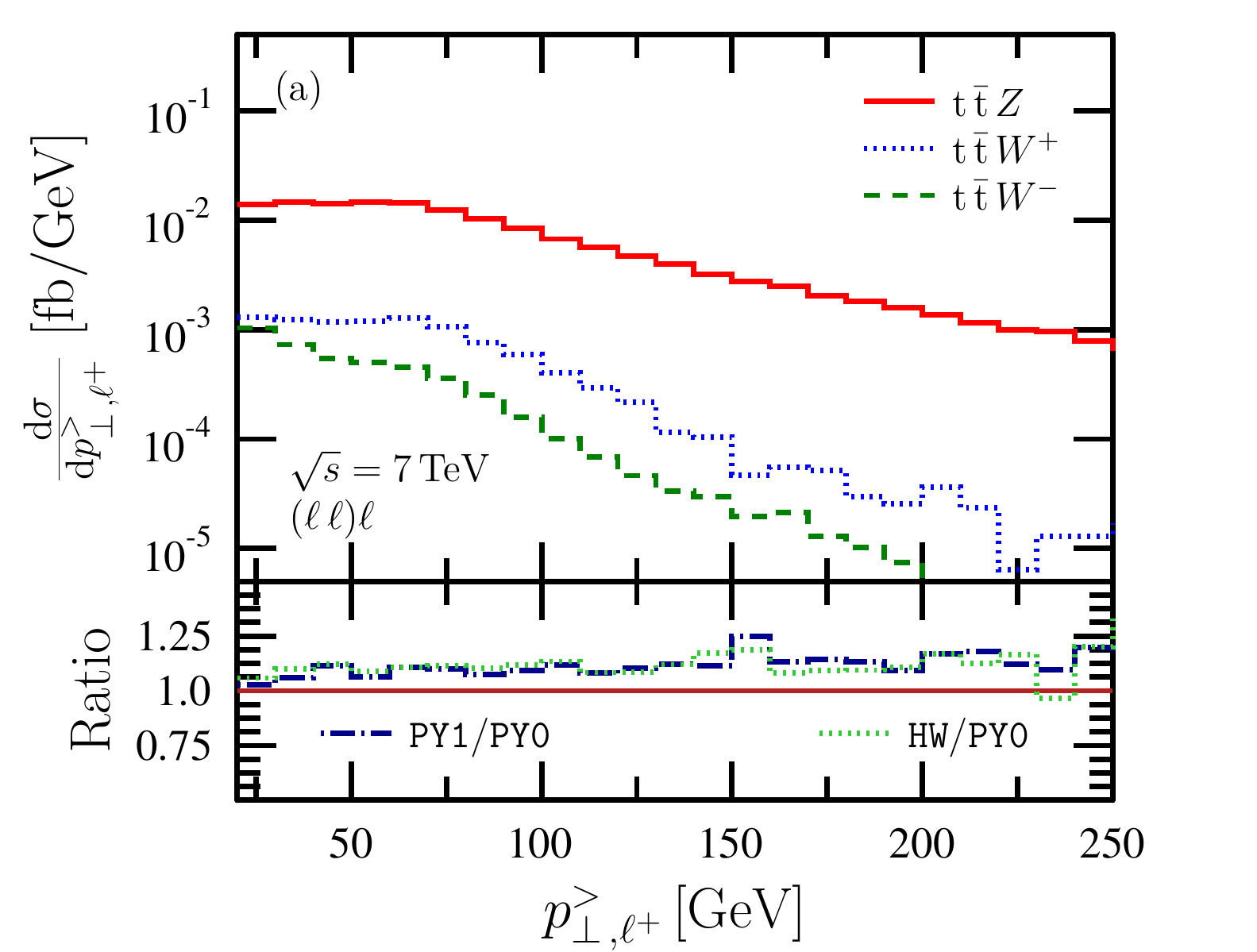}
\includegraphics[width=0.49\textwidth]{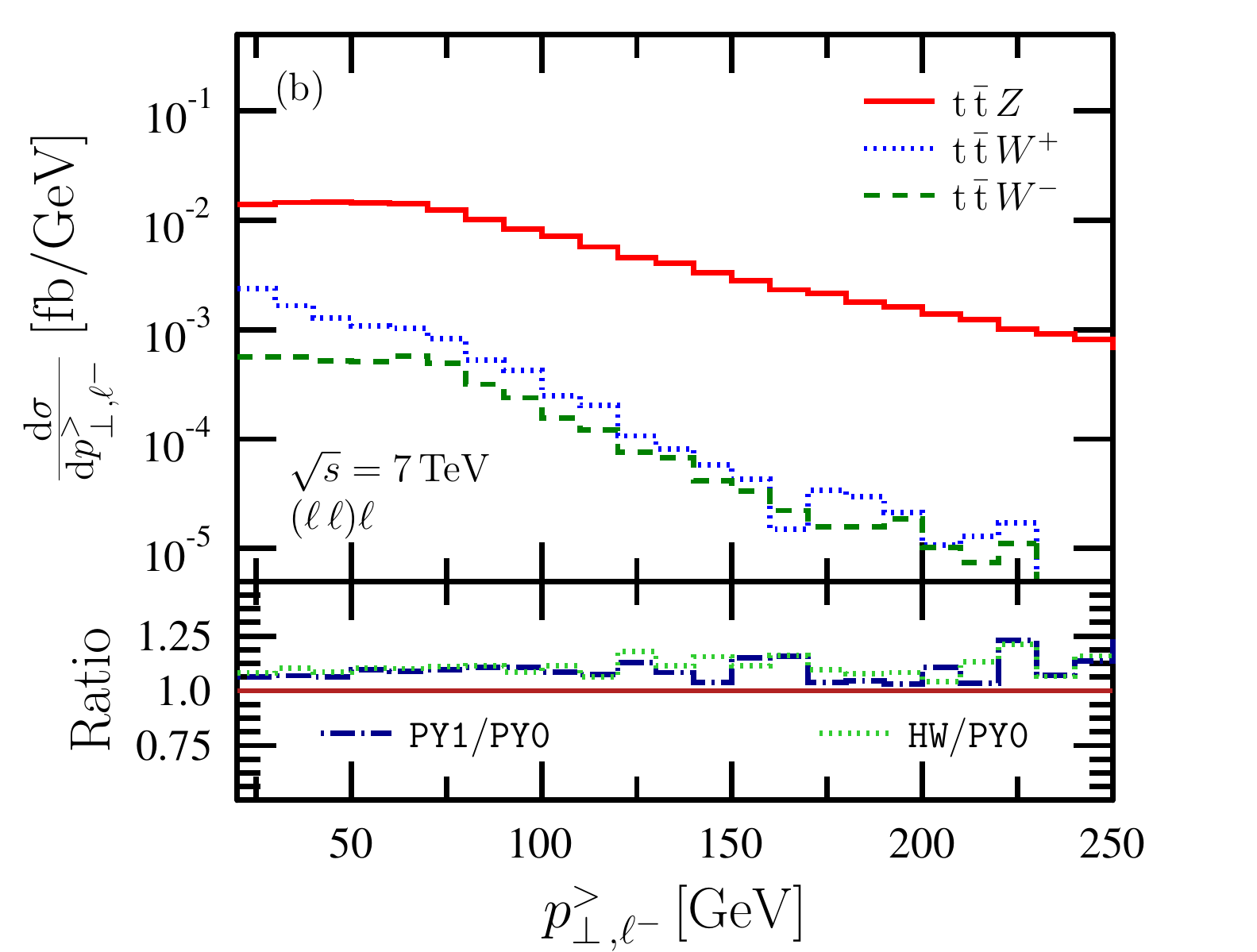}
\caption{\label{ptlepantilep} Transverse momentum distributions of
a) the hardest anti-lepton and
b) the hardest lepton of each event, 
at $\sqrt{s} = 7$\,TeV LHC, as predicted by \powhel\ +~\pythia\ after
the trilepton analysis for the processes \ttz\ (solid), \ttwp\ (dotted)
and \ttwm\ (dashed).
In the lower inset the ratios between cumulative results using different
SMC \herwig\ and \pythia\ (\hw/\py0) and between cumulative results
obtained by neglecting and including photon bremsstrahlung from leptons
in \pythia\ (\py1/\py0) are also shown.
}
\end{figure}

It is also interesting to separate the behaviour of leptons and
anti-leptons.  As seen in Fig.~\ref{ptlepantilep}.a, the
$p_\bot$-distribution of the hardest lepton of each event has a plateau
in the region 20--70 GeV in case of both \ttz\ and \ttwp, whereas for
the distribution of the hardest anti-lepton plotted in 
Fig.~\ref{ptlepantilep}.b, the plateau appears for \ttz\ and \ttwm,
i.e.~the situation for \ttwp\ and \ttwm\ is symmetric.  This symmetry
suggests that these plateaus are generated by the selection cuts and
include the contributions of the prompt leptons and anti-leptons
originated directly from the decay of the initial $Z$ and $W$
weak-bosons (whereas possible secondary leptons or anti-leptons with
the same transverse momentum are cut).  The behaviour of the tails of
the \ttwp\ and \ttwm\ distributions Fig.~\ref{ptlepantilep}.b is the
same as already observed in Fig.~\ref{ptmissnocut}.a for the case of
the inclusive analysis (see discussion in Subsection~\ref{inclu}),
suggesting that even the high-$p_\bot$ regions are dominated by prompt
leptons, as expected.
In case of both \herwig\ and \pythia\ where the photon bremsstrahlung
effect is switched off, these distributions are just rescaled by a
5--10\,\% factor, slightly increasing towards the tail of the
distributions, indicating that hardest (anti-)leptons are more prone to
emit photons than the softer ones. 

We also repeated the analysis in the trilepton channel in case of an
LHC $\sqrt{s} = 8$\,TeV center-of-mass energy, that can be useful in
view of future data analysis on the basis of the events recorded in the
present run. For future reference, we report here our cumulative
predictions for the cross-section contributions of the three processes
\ttz, \ttwp\ and \ttwm\ at $\sqrt{s} = 8$\,TeV:
$\sigma_{(e,e),e} = 0.782$\,fb,
$\sigma_{(e,e),\mu} = 0.388$\,fb,      
$\sigma_{(\mu,\mu),e} = 0.420$\,fb,
$\sigma_{(\mu,\mu),\mu} = 0.934$\,fb,
$\sigma_{{\scriptscriptstyle \sum}} = 2.524$\,fb,
all with a statistical uncertainty below $5 \cdot 10^{-5}$\,fb. 
Furthermore, predictions for the same differential distributions
already discussed in the $\sqrt{s} = 7$\,TeV case, were produced in the
8\,TeV case, and we have found very similar results, except for a
rescaling factor just given by the ratio of the cross-sections at 8 and
7\,TeV. The LHE's are freely available at our web repository:
{\texttt {http://www.grid.kfki.hu/twiki/bin/view/DbTheory/WebHome}}.

\subsection{Dilepton-channel analysis}
\label{dilepton}

As mentioned in the introduction, studies of \ttv\ decays in the
dilepton channel, with two same-sign leptons plus jets, have their
original motivation that this kind of signature is hardly produced by
SM processes, and can thus be used in searches for supersymmetry. In this
case, \ttv\ can be considered as a background with respect to possible
new physics processes. Other sizable backgrounds involve many different
diboson and triboson production processes. An exhaustive list of
backgrounds in this context can be found in Ref.~\cite{Campbell:2012dh}.
New physics searches usually also involve  a cut on missing energy.
In this paper we explore the dilepton channel, without imposing any
missing energy cut, as also done in the very recent CMS technical
report~\cite{CMSnew}, where the analysis was optimized on the basis of
data collected at LHC at $\sqrt{s} = 7$\,TeV corresponding to an
integrated luminosity $L = 4.98\,\rm{fb}^{-1}$ . This way the
relatively small number of \ttv\ events does not suffer any further
suppression due to this cut.

The aim of this analysis is to select the events where one of the
quarks of the \tT-pair decays leptonically and the other one
hadronically, and the vector boson decays leptonically giving rise to a
lepton with the same sign of the lepton coming from the (anti-)quark.
In case of \ttwp-production this means that we are looking for
$W^+\to \ell^+\,\nu_\ell$, accompanied by the leptonic decay of the
t-quark, whereas, in case of \ttwm-production we aim to select events
with $W^- \to \ell^-\,\bar{\nu}_\ell$, accompanied by the leptonic
decay of the $\bar{\rm{t}}$-quark. In case of \ttz-production
$Z\to \ell^+ \ell^-$, and thus it is sufficient that either the  t- or
the $\bar{\rm t}$-quark decays leptonically. 

Following the CMS Collaboration, we considered the following set of cuts:
\begin{enumerate}
\item
two same-sign isolated leptons with $p_{\bot, \, {\ell_1}} > 55$\,GeV 
and $p_{\bot, \, {\ell_2}} > 30$\,GeV, respectively, within CMS
acceptance ($|\eta_\ell| < 2.4$, plus a further removal of the
[1.4442, 1.566] pseudorapidity range corresponding to the ECAL
barrel/endcap transition region, applied in case of electrons),
\item
dilepton invariant mass $m_{\ell_1,\ell_2} >$ 8\,GeV,
\item
at least 3 jets with $p_{\bot, \, j} > 20$\,GeV  and $|\eta_j| < 2.4$, 
satisfying the additional cut $\Delta R (j,\ell) > 0.4$
on the distance in the pseudorapidity-azimuthal angle plane, 
for both $\ell = \ell_1, \ell_2$, 
\item
at least one of the previous 3 jets must be b-tagged,
\item
$H_{\rm T} > 100$\,GeV, where $H_{\rm T}$ is computed as the scalar sum
of the transvers momenta of all jets satisfying cut 3).
\end{enumerate}

Jets were constructed using of the anti-$\rm{k_\bot}$ algorithm, 
with $R = 0.5$, as implemented in \fastjet~\cite{Cacciari:2011ma}. 
Lepton isolation was computed by making use of the standard 
isolation criterion mentioned in the CMS technical report~\cite{CMSsusy}:
we require a lepton relative isolation $I_{rel} > 0.15$, where
$I_{rel}$ is computed as the ratio between the scalar sum of the
transverse momenta of all tracks within a distance $\Delta R < 0.3$
with respect to the selected lepton and the transverse momentum of the
lepton itself (excluded from the sum at the numerator).  Furthermore,
in case of multiple dilepton pairs satisfying the cuts mentioned above,
the pair was selected with the largest combined transverse momentum. 

As also done in the CMS analysis~\cite{CMSnew}, we explicitly exclude
from this analysis all events that are selected in the trilepton
channel analysis, in order to obtain two statistical independent
samples (trilepton veto). As we will see in the following, the final
predictions in the dilepton channel as for both the number of events
and the shape of the distributions, will indeed be affected by this
choice, especially as for the \ttz\ process.   

\begin{wrapfigure}{r}{0.6\textwidth}
\begin{center}
\includegraphics[width=0.6\textwidth]{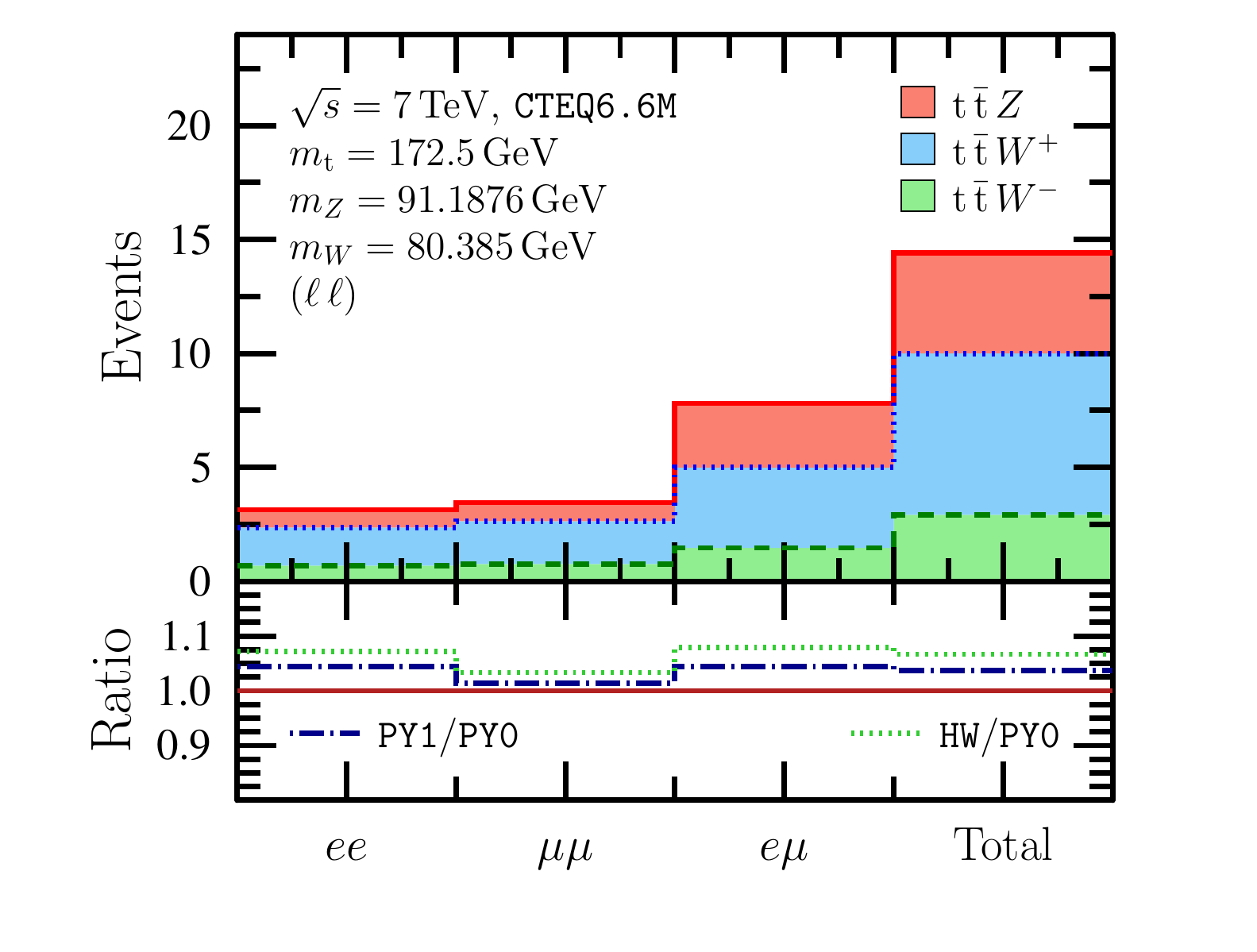}
\caption{\label{figeventsdil}
Number of events in the dilepton channel at $\sqrt{s} = 7$\,TeV LHC, as
predicted by \powhel\ +~\pythia, for an integrated luminosity 
$L = 4.98$\,fb$^{-1}$. The contribution in the ($e$, $e$), ($\mu$,
$\mu$), ($e$, $\mu$) channels are shown separately, as well as their
sum in the last bin.  The contributions due to \ttz, \ttwp\ and \ttwm\
are cumulated one over the other.
In the lower inset the ratios between cumulative results using different
SMC \herwig\ and \pythia\ (\hw/\py0) and between cumulative results
obtained by neglecting and including photon bremsstrahlung from leptons
in \pythia\ (\py1/\py0) are also shown.
}
\end{center}
\end{wrapfigure}

Differences between our theoretical analysis framework and the 
experimental conditions are listed in the following:
\begin{itemize}
\item
Contrary to experimental reconstruction of the events, electron and
muon detection efficiencies in our theoretical simulations were assumed
to be 100\,\% and charge misidentification effects neglected.
\item
Also, while in the experiment b-jets were reconstructed as displaced
vertices, making use of spatial tracking information, and a b-tagging
algorithm was applied, ensuring a limited efficiency in the
reconstruction of b-jets, accompanied by a non-negligible fake rate, in
our simulations we identified b-jets using the \texttt{MCTRUTH}
parameter which allows for tracking back b and $\bar{\rm b}$ quarks
from \tT-decay, but we lacked spatial information concerning the position
of displaced vertices.
\end{itemize}
Despite the differences in the analysis, and perhaps other experimental
detail we are not aware of, the theoretical predictions, shown in
Fig.~\ref{figeventsdil}, are compatible with the experimental results.

The largest contribution to the total number of events is from the
\ttwp\ process, followed by the \ttz\ and the \ttwm\ ones.  The
contribution of the \ttwp\ process is larger than the \ttwm\ one
already at the inclusive level (see \sect{trilepton}),
with the ratio between the two remaining
almost the same after cuts (2.45 for the inclusive predictions and 2.42
after cuts). The contribution of \ttwp\ is enhanced with respect to
that of \ttz\ after cuts is an effect of the selection cuts and of the
trilepton veto. For the \ttw\ processes, the contribution in the
($e$, $\mu$) channel turns out to be almost twice the average of the
($e$, $e$) and ($\mu$, $\mu$) ones, as naively expected on the basis of
the possible charge and flavour combinations. (An electron can come from
the W and a muon with the same sign from one of the t-quarks, or
viceversa.) For the \ttz\ process, the contribution in the ($e$, $\mu$)
channel turns out to be $\simeq 3.5$ times the average of the ($e$, $e$)
and ($\mu$, $\mu$) ones, i.e.~larger than expected on the basis of the
charge and flavour combinatorics. The reason has to be attributed
to the trilepton veto. As seen in Fig.~\ref{figeventstri}, the number
of events in the trilepton channel in case of the ($e$, $e$) $e$ and
($\mu$, $\mu$) $\mu$  combinations are larger than those in the
($e$, $e$)~$\mu$ and ($\mu$, $\mu$)~$e$ bins. The former affect
the ($e$, $e$) and ($\mu$, $\mu$) bins of the dilepton analysis,
while the latter affect the ($e$, $\mu$) bins of the dilepton analysis.
As a consequence, the contribution to the ($e$, $\mu$) channel of the
dilepton analysis is less suppressed than those in the ($e$, $e$) and
($\mu$, $\mu$) channels due to the trilepton veto.  The predictions by
different SMC programs, i.e. \herwig\ and \pythia\ where the photon
bremsstrahlung from leptons is switched off, are up to 8\,\% and 5\,\%
larger than those of \pythia, as can be seen from the lower panel of
Fig.~\ref{figeventsdil}. These differences have the same sign,
but are smaller, than those found in case of the trilepton analysis
(see the lower panel of Fig.~\ref{figeventstri} for comparison).  

The \powhel~+~\pythia\ predictions for the cross-section contributions
in the different dilepton channels (see Fig.~\ref{figeventsdil}),
summing over \ttz, \ttwp\  and \ttwm, in case of $\sqrt{s} = 7$\,TeV LHC
are listed in the following, together with their sum:
$\sigma_{(e,e)} = 0.631$\,fb,
$\sigma_{(e,\mu)} = 0.694$\,fb,      
$\sigma_{(\mu,\mu)} = 1.569$\,fb,
$\sigma_{{\scriptscriptstyle \sum}} = 2.894$\,fb,
all with a statistical uncertainty below $3 \cdot 10^{-5}$\,fb.

\begin{figure}[h!]
\includegraphics[width=0.49\textwidth]{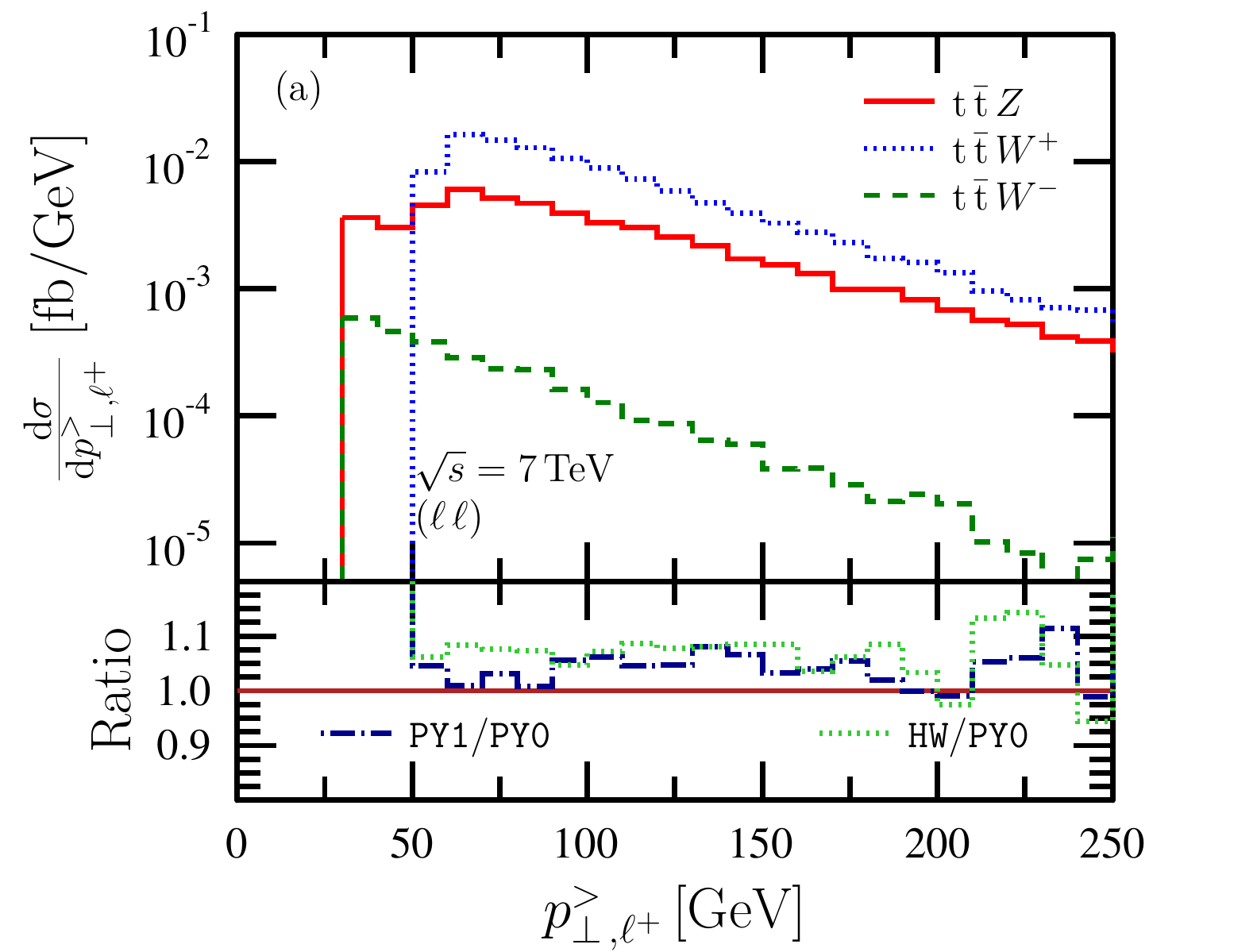}
\includegraphics[width=0.49\textwidth]{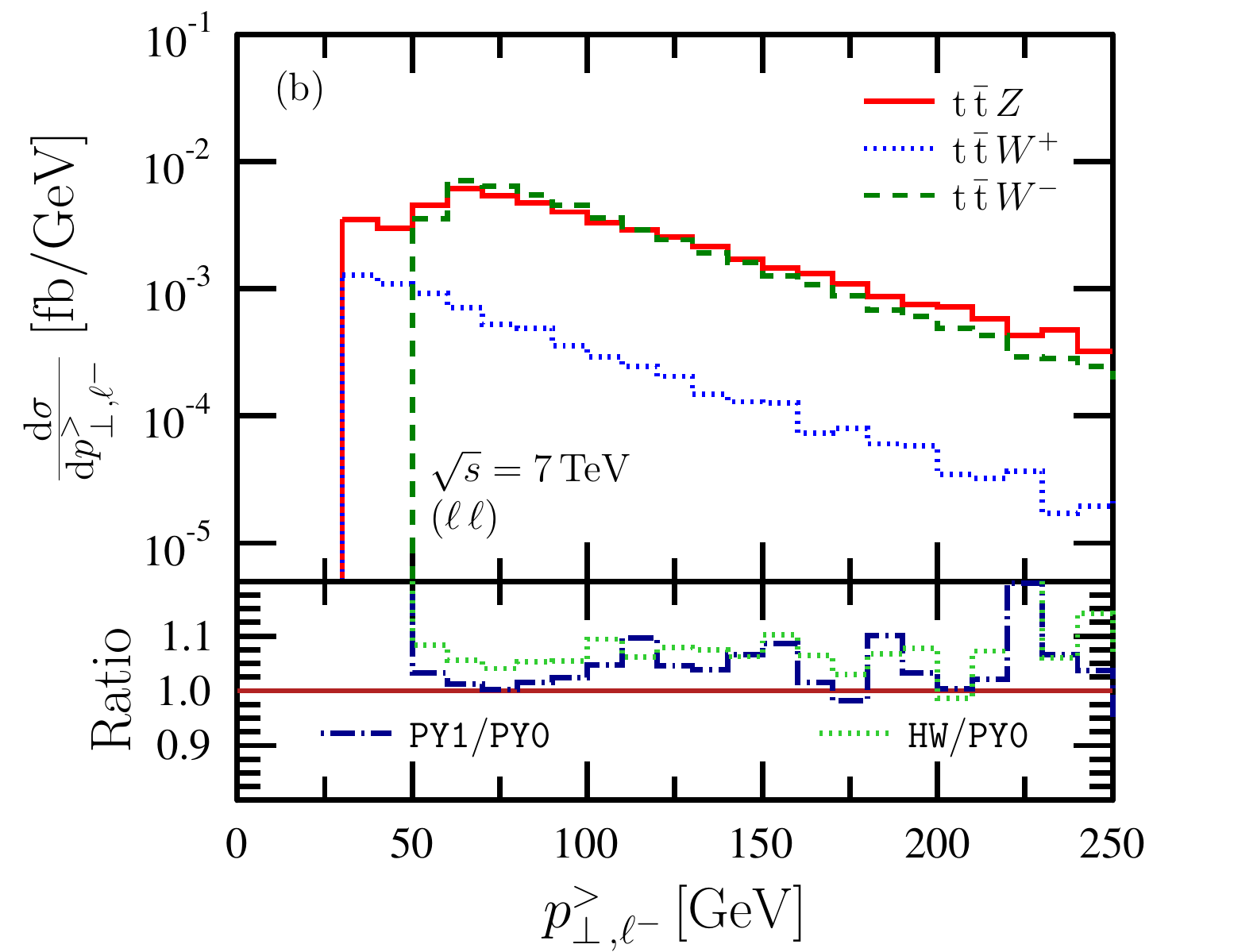}
\caption{\label{pthardestlep} Transverse momentum distributions of
a) the hardest anti-lepton and
b) the hardest lepton of each event  
at $\sqrt{s} = 7$\,TeV LHC, as predicted by \powhel\ +~\pythia\ after
the dilepton analysis.  The distributions for \ttz, \ttwp\ and \ttwm\
are shown by solid (red), dotted (blue) and dashed (green) lines,
respectively. 
In the lower inset the ratios between cumulative results using different
SMC \herwig\ and \pythia\ (\hw/\py0) and between cumulative results
obtained by neglecting and including photon bremsstrahlung from leptons
in \pythia\ (\py1/\py0) are also shown.
}
\end{figure}

As for the comparison with the experimental data, we note that in the
CMS technical report~\cite{CMSnew} a contribution to the number of
events was assigned to the effect of charge misidentification for the
leptons, in particular the electrons, and another additional
contribution to the effect of non-prompt leptons, i.e.~leptons not
coming directly from heavy bosons decays.
In our theoretical simulations the background due to charge
misidentification vanishes, whereas a possible contribution of
non-prompt leptons to our final results relies on the effectiveness of
the isolation criteria we adopted.  In this respect, even if we lack a
precise estimate, it can be interesting to observe the differential
distributions of the hardest isolated (anti-)leptons of each event
after cuts, plotted in Fig.~\ref{pthardestlep}. 

We see from Fig.~\ref{pthardestlep}.a, in case of \ttwp\ the hardest
isolated anti-lepton after cuts has a minimum $p_\bot$ of 50\,GeV and a
peak slightly above it, whereas in case of \ttwm\ it has a minimum
$p_\bot$ of 30\,GeV without a peak. In case of the hardest isolated
lepton, instead, the behaviour of \ttwp\ and \ttwm\ is the opposite,
as can be seen in Fig.~\ref{pthardestlep}.b.  This behaviour is
compatible with cut 1) and means that the system of proposed cuts is
effective in selecting prompt leptons, i.e. the selection of
($\ell^+$, $\ell^+$) pair in case of \ttwm\ decay, or of ($\ell^-$,
$\ell^-$) pair in the \ttwp\ decay are actually suppressed
by orders of magnitude, even if several leptons and anti-leptons can be
present after PS, hadronization and hadron decays.  In case of \ttz\
decays, two opposite-charge leptons are produced by $Z$-decays, so
both ($\ell^+$,~$\ell^+$) and ($\ell^-$, $\ell^-$) pairs of prompt
leptons could be selected. Thus a peak above 50\,GeV is present in both
the lepton and the anti-lepton distributions.  In all cases, the peaks
slightly above 50\,GeV are related to the request of having at least
one (anti-)lepton with $p_\bot > 55$\,GeV in the selection cuts. 

\begin{figure}[h!]
\begin{center}
\includegraphics[width=0.49\textwidth]{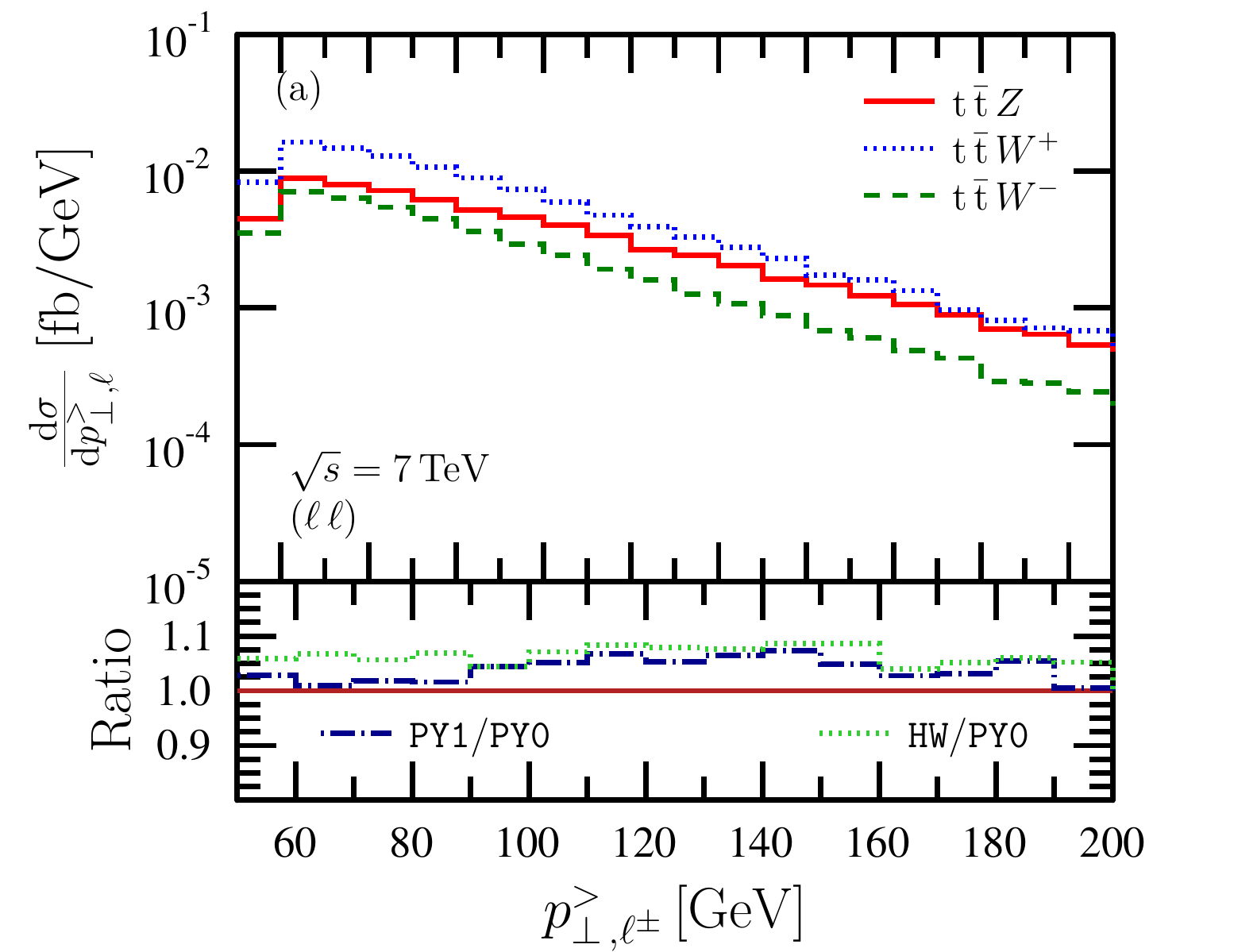}
\includegraphics[width=0.49\textwidth]{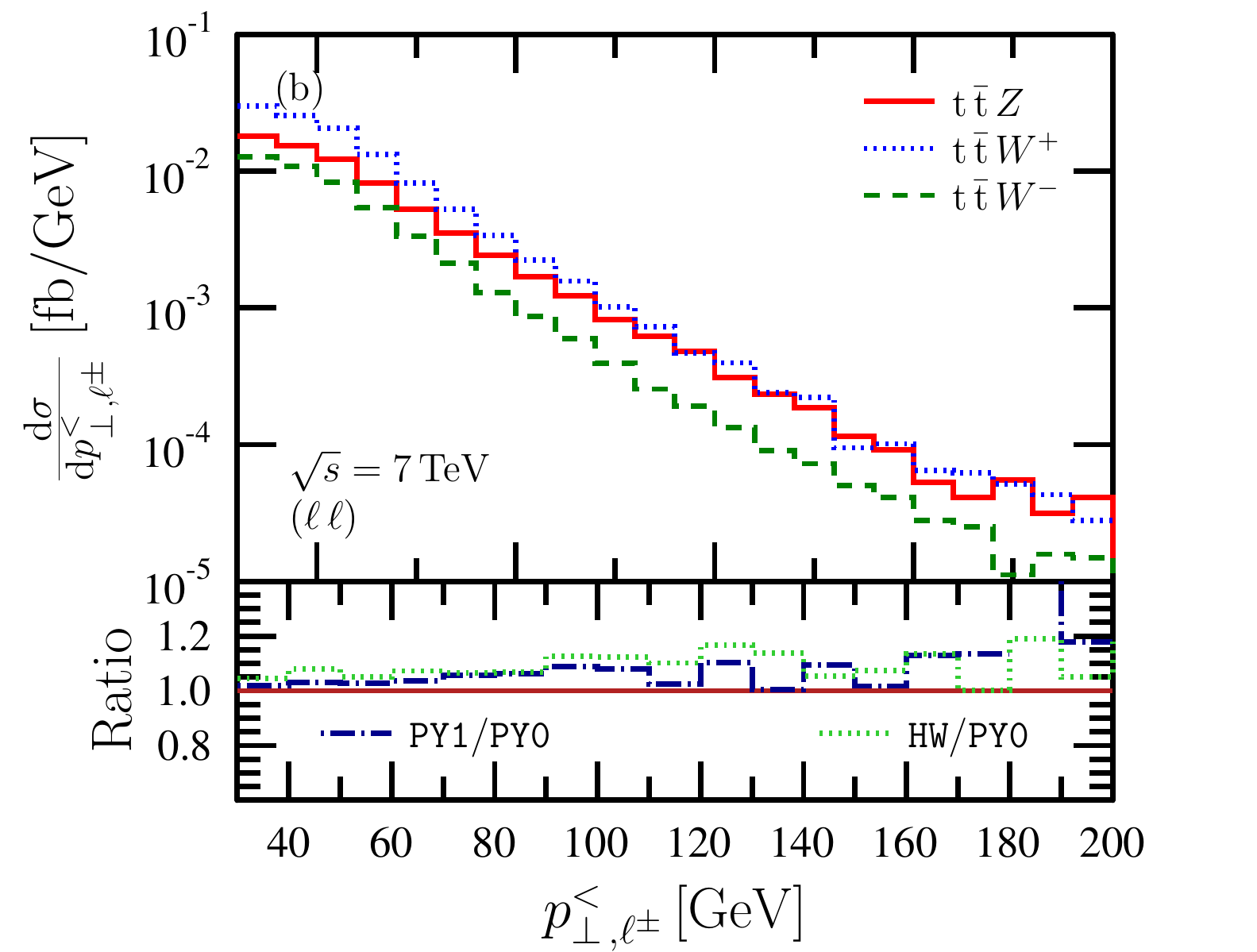}
\caption{\label{leadingpt} Transverse momentum distribution of
a) the leading and
b) the subleading (anti)-lepton
of each same-sign ($\ell$, $\ell$) pair after the dilepton analysis. 
Predictions by \powhel\ +~\pythia\ at $\sqrt{s} = 7$\,TeV LHC
corresponding to the different \ttz, \ttwp and \ttwm\ processes are
shown separately.
In the lower inset the ratios between cumulative results using different
SMC \herwig\ and \pythia\ (\hw/\py0) and between cumulative results
obtained by neglecting and including photon bremsstrahlung from leptons
in \pythia\ (\py1/\py0) are also shown.
}
\end{center}
\end{figure}

As examples of further distributions that can be measured in the
experiment, the cumulative transverse momentum distributions of the 
leading and subleading lepton or anti-lepton of the ($\ell$, $\ell$) 
selected pairs are plotted in Fig.~\ref{leadingpt}.  At low $p_\bot$
the sum is dominated by the \ttwp\ contribution, whereas in the high
$p_\bot$ tail (i.e.~above $\simeq 300$\,GeV in case of the leading lepton
and above $\simeq 150$\,GeV in case of the subleading one), the
contributions of \ttz\ and \ttwp\ become almost equal.    

\begin{figure}[h!]
\begin{center}
\includegraphics[width=0.49\textwidth]{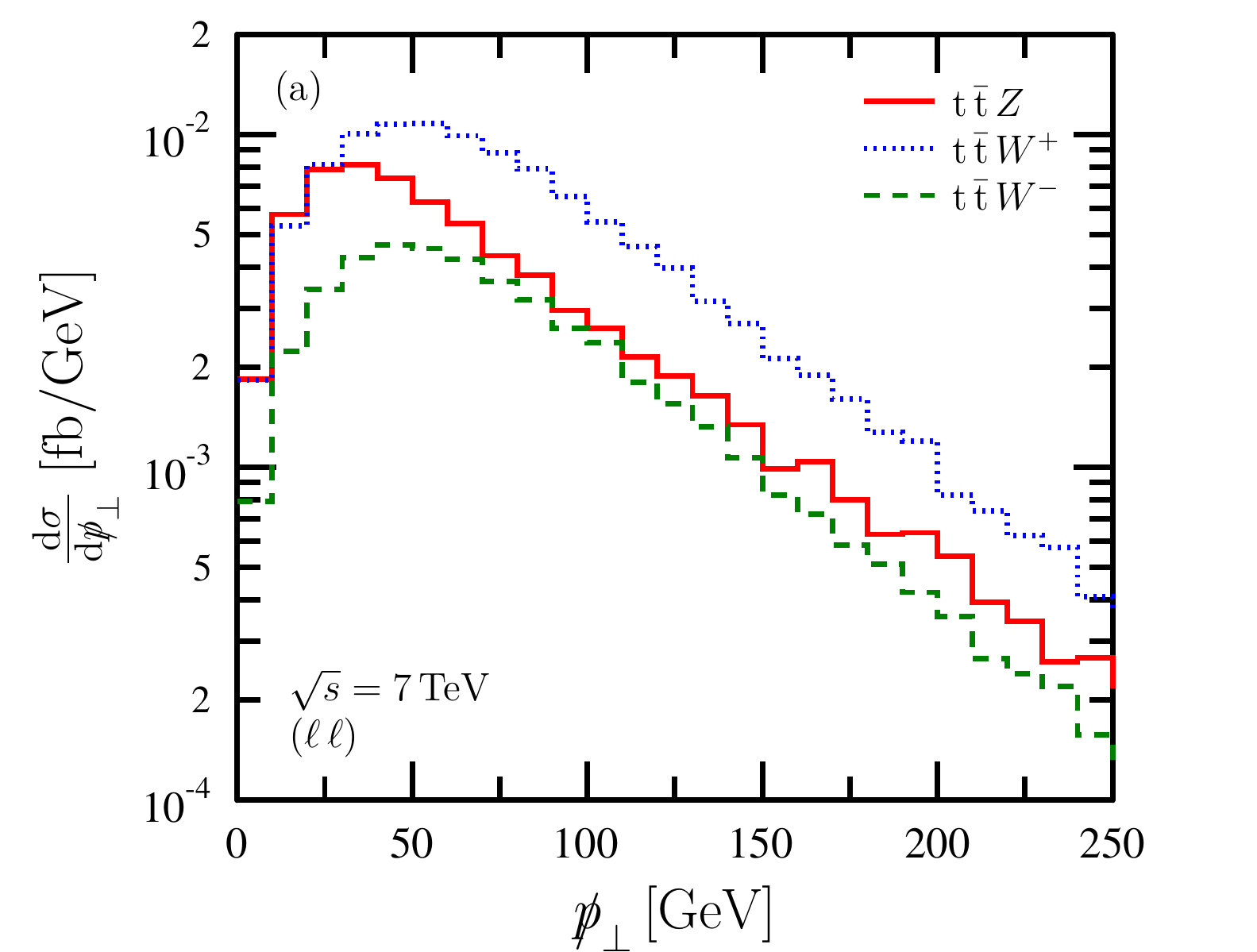}
\includegraphics[width=0.49\textwidth]{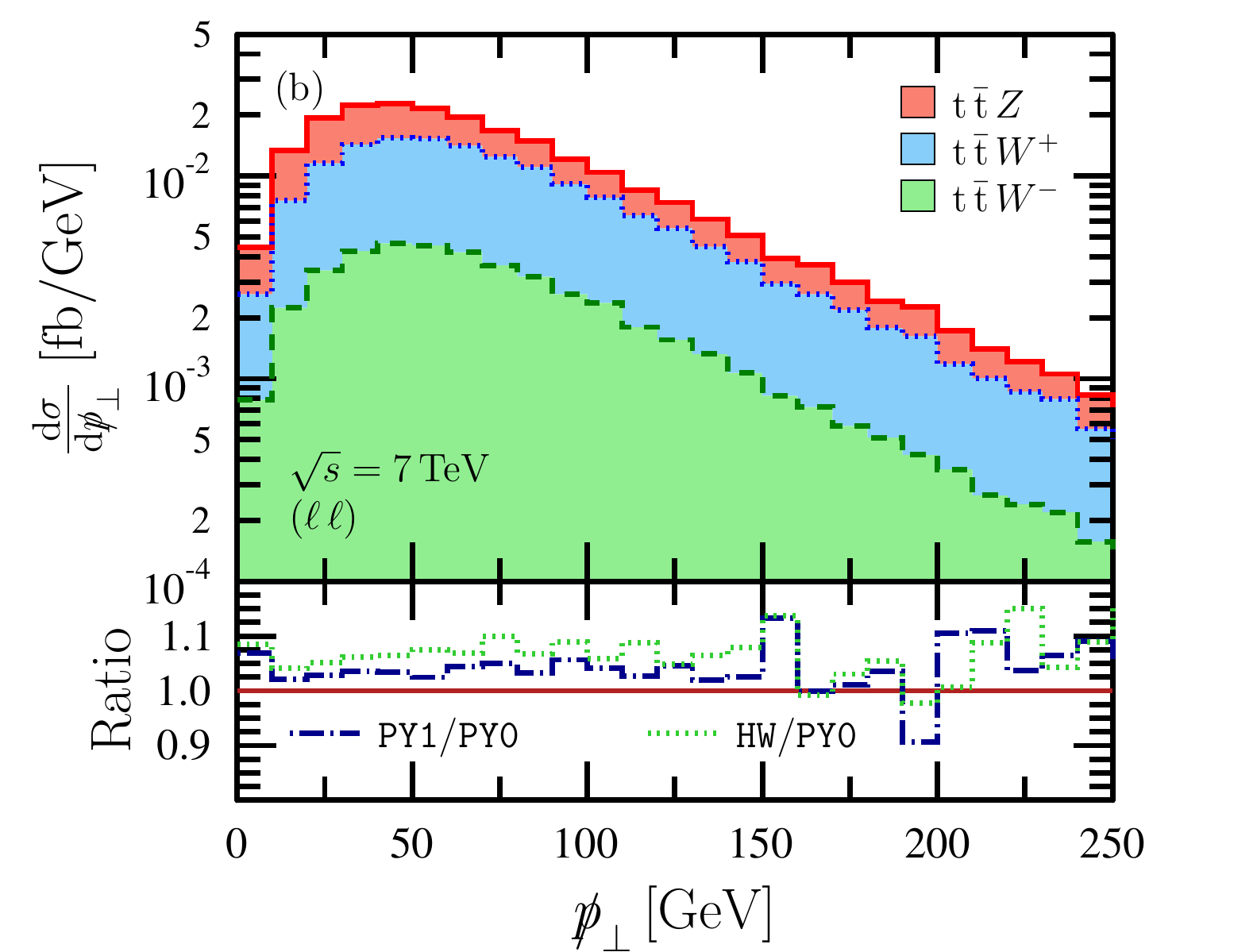}
\caption{\label{missing} Missing transverse momentum distribution 
at $\sqrt{s} = 7$\,TeV LHC, as predicted by \powhel\ +~\pythia\ after the
dilepton analysis.
a) distributions for the processes \ttz\ (red), \ttwp\ (dotted)
and \ttwm\ (dashed) (red), dotted (blue) and slashed (green) lines.
b) these different contributions are added one over the other in a
cumulative way. 
In the lower inset the ratios between cumulative results using different
SMC \herwig\ and \pythia\ (\hw/\py0) and between cumulative results
obtained by neglecting and including photon bremsstrahlung from leptons
in \pythia\ (\py1/\py0) are also shown.
}
\end{center}
\end{figure}

In view of the searches for new physics in the dilepton channel
another interesting distribution is that of the missing transverse
energy, plotted in Fig.~\ref{missing}. In Fig.~\ref{missing}.a, 
different shapes characterize the three \ttv\ processes. The
distribution for \ttz\ is peaked around 30\,GeV, while that for \ttw\
is peaked around 50\,GeV. This difference is related to the
$W \to \ell \, \nu_\ell$ decay events selected in the dilepton analysis, 
that populate the peak region. The suppression in the first few bins,
not present in the analogous inclusive \pTmiss-distribution plotted in
Fig.~\ref{ptmissnocut}.b, is an effect of the set of cuts, aiming at
the selection of two same-sign prompt (anti-)leptons. With this selection
both the primary boson and either the t- or the $\bar{t}$-quark should
decay leptonically, leading to a non-zero \pTmiss. As expected,
including further cuts on \pTmiss, will enhance the relative
contribution of the \ttw\ process with respect to the \ttz\ one, and
will reduce the number of observed \ttv\ events.  In particular,
integrating over the cumulative \pTmiss-distribution, plotted in
Fig.~\ref{missing}.b, we find that a cut of $\pTmiss > 50$\,GeV
corresponds to a reduction on the total number of events, plotted in
Fig.~\ref{figeventsdil}, by a factor of $\simeq 4$ and a cut of
$\pTmiss > 100$\,GeV to a further reduction by a similar factor.  

Looking forward to an analysis of data collected in the recent LHC
energy upgrade, we repeated the whole analysis at $\sqrt{s} = 8$\,TeV
LHC.  For future reference, we list our predictions for the
cross-sections after cuts at this energy for each dilepton channel,
together with their sum. We found
$\sigma_{(e,e)} = 0.907$\,fb,
$\sigma_{(e,\mu)} = 0.991$\,fb,      
$\sigma_{(\mu,\mu)} = 2.289$\,fb,
$\sigma_{{\scriptscriptstyle \sum}} = 4.187$\,fb,
all with a statistical uncertainty $<$ 5 $\cdot$ 10$^{-5}$.
As for differential distributions at 8\,TeV, we found that their general
qualitative behaviour and their shapes are similar to those already
shown at 7\,TeV, thus we do not present them again here.  These can
just be obtained  by a proper rescaling factor given by the ratio of
the cross-sections at 8 and 7\,TeV. The LHE's are freely available at
our web repository.  

\section{Conclusions}
\label{conclusions}

In this paper we provided predictions for \ttv\ hadroproduction at LHC.
Our NLO predictions in the \ttw\ channel provide a confirmation,
by a completely independent method, of those already presented by
Ref.~\cite{Campbell:2012dh} at 7, 8 and 14\,TeV center-of-mass
energies. In case of both \ttz\ and \ttw\, we quote the uncertainties
on the cross-sections due to scale variation at NLO.  Furthermore, we
provide predictions for differential distributions due to these same
processes at the hadron level, thanks to the matching of the NLO
computation with a SMC approach, through the POWHEG method as
implemented in the \powhel\ framework, on the basis of the interface of
the \powhegbox\ and the \helacnlo\ event generators.  In case of \ttw\
these are the first predictions provided at such a level of accuracy
in the literature, whereas in case of \ttz\ this paper can be
considered a sequel to our previous ones, exploring different decay
channels in the phenomenological analysis. In particular, we
concentrate on the dilepton and trilepton channels, also recently
studied by the CMS Collaboration.

At 7\,TeV our predictions turn out to be compatible with the
experimental data, taking into account the slightly different
selection conditions, whereas, in case of 8\,TeV, where the
experimental data have not yet been analyzed, we provide new
predictions, both for the total and differential cross-sections, that
we hope can be useful for the experimental analysis. For this purpose
all LHE files with sets of several millions of events at the first
radiation emission level used for the analysis presented in this paper,
are freely available at our web repository:
{\texttt {http://www.grid.kfki.hu/twiki/bin/view/DbTheory/WebHome}}.

In the future, when the accumulated data will reach the necessary
statistics that allows for a more detailed comparison of the theory
with the experiment, the inclusion of spin correlations effects and
radiative corrections in top quark and heavy bosons decays, as well as
the evaluation of the different backgrounds at the same level of
accuracy, will be possible and indispensable. Moreover the estimate of
the theoretical uncertainty will require further studies of the scale
variation at the NLO and SMC levels, as well as a systematic study of the
effect of different PDF sets and parameters.

\section{Acknowledgements}
This research was supported by
the LHCPhenoNet network PITN-GA-2010-264564,
the Swiss National Science Foundation Joint Research Project SCOPES
IZ73Z0\_1/28079, the T\'AMOP 4.2.1./B-09/1/KONV-2010-0007
and 4.2.2/B-10/1-2010-0024 projects, the
Hungarian Scientific Research Fund grant K-101482.
We are grateful to G.~Dissertori, F.~Pandolfi and P.~Skands for useful discussions.  



\begin{thebibliography}{10}

\bibitem{CMSsusy}
The CMS Collaboration, {\it Search for new physics with same-sign isolated
  dilepton events with jets and missing energy},  Tech. Rep. CMS PAS
  SUS-11-010, July, 2011.
\newblock Available on the CERN CDS information server.

\bibitem{CMSnew}
The CMS Collaboration, {\it First measurement of vector boson production
  associated with top-antitop pairs at 7 tev},  Tech. Rep. CMS PAS TOP-12-014,
  July, 2012.
\newblock Available on the CERN CDS information server.

\bibitem{Alioli:2010xd}
S.~Alioli, P.~Nason, C.~Oleari, and E.~Re, {\it {A general framework for
  implementing NLO calculations in shower Monte Carlo programs: the POWHEG
  BOX}},  {\em JHEP} {\bf 06} (2010) 043,
  [\href{http://xxx.lanl.gov/abs/1002.2581}{{\tt arXiv:1002.2581}}].

\bibitem{Frixione:2007vw}
S.~Frixione, P.~Nason, and C.~Oleari, {\it {Matching NLO QCD computations with
  Parton Shower simulations: the POWHEG method}},  {\em JHEP} {\bf 11} (2007)
  070, [\href{http://xxx.lanl.gov/abs/0709.2092}{{\tt arXiv:0709.2092}}].

\bibitem{Nason:2004rx}
P.~Nason, {\it {A New method for combining NLO QCD with shower Monte Carlo
  algorithms}},  {\em JHEP} {\bf 0411} (2004) 040,
  [\href{http://xxx.lanl.gov/abs/hep-ph/0409146}{{\tt hep-ph/0409146}}].

\bibitem{Bevilacqua:2011xh}
G.~Bevilacqua, M.~Czakon, M.~Garzelli, A.~van Hameren, A.~Kardos, et~al., {\it
  {HELAC-NLO}},  \href{http://xxx.lanl.gov/abs/1110.1499}{{\tt
  arXiv:1110.1499}}.

\bibitem{Kardos:2011qa}
A.~Kardos, C.~Papadopoulos, and Z.~Tr\'ocs\'anyi, {\it {Top quark pair production
  in association with a jet with NLO parton showering}},  {\em Phys.Lett.} {\bf
  B705} (2011) 76--81, [\href{http://xxx.lanl.gov/abs/1101.2672}{{\tt
  arXiv:1101.2672}}].

\bibitem{Garzelli:2011vp}
M.~Garzelli, A.~Kardos, C.~Papadopoulos, and Z.~Tr\'ocs\'anyi, {\it {Standard Model
  Higgs boson production in association with a top anti-top pair at NLO with
  parton showering}},  {\em Europhys.Lett.} {\bf 96} (2011) 11001,
  [\href{http://xxx.lanl.gov/abs/1108.0387}{{\tt arXiv:1108.0387}}].

\bibitem{Garzelli:2011iu}
M.~Garzelli, A.~Kardos, and Z.~Tr\'ocs\'anyi, {\it {NLO event samples for the
  LHC}},  \href{http://xxx.lanl.gov/abs/1111.1446}{{\tt arXiv:1111.1446}}.

\bibitem{Dittmaier:2012vm}
S.~Dittmaier, S.~Dittmaier, C.~Mariotti, G.~Passarino, R.~Tanaka, et~al., {\it
  {Handbook of LHC Higgs Cross Sections: 2. Differential Distributions}},
  \href{http://xxx.lanl.gov/abs/1201.3084}{{\tt arXiv:1201.3084}}.

\bibitem{Kardos:2011na}
A.~Kardos, Z.~Tr\'ocs\'anyi, and C.~Papadopoulos, {\it {Top quark pair production
  in association with a Z-boson at NLO accuracy}},  {\em Phys.Rev.} {\bf D85}
  (2012) 054015, [\href{http://xxx.lanl.gov/abs/1111.0610}{{\tt
  arXiv:1111.0610}}].

\bibitem{Garzelli:2011is}
M.~Garzelli, A.~Kardos, C.~Papadopoulos, and Z.~Tr\'ocs\'anyi, {\it {Z0 - boson
  production in association with a top anti-top pair at NLO accuracy with
  parton shower effects}},  {\em Phys.Rev.} {\bf D85} (2012) 074022,
  [\href{http://xxx.lanl.gov/abs/1111.1444}{{\tt arXiv:1111.1444}}].

\bibitem{Campbell:2012dh}
J.~M. Campbell and R.~K. Ellis, {\it {$t \bar{t} W^{+-}$ production and decay
  at NLO}},  {\em JHEP} {\bf 1207} (2012) 052,
  [\href{http://xxx.lanl.gov/abs/1204.5678}{{\tt arXiv:1204.5678}}].

\bibitem{Sjostrand:2006za}
T.~Sjostrand, S.~Mrenna, and P.~Z. Skands, {\it {PYTHIA 6.4 Physics and
  Manual}},  {\em JHEP} {\bf 0605} (2006) 026,
  [\href{http://xxx.lanl.gov/abs/hep-ph/0603175}{{\tt hep-ph/0603175}}].

\bibitem{Corcella:2002jc}
G.~Corcella, I.~Knowles, G.~Marchesini, S.~Moretti, K.~Odagiri, et~al., {\it
  {HERWIG 6.5 release note}},
  \href{http://xxx.lanl.gov/abs/hep-ph/0210213}{{\tt hep-ph/0210213}}.

\bibitem{vanHameren:2009dr}
A.~van Hameren, C.~Papadopoulos, and R.~Pittau, {\it {Automated one-loop
  calculations: A Proof of concept}},  {\em JHEP} {\bf 0909} (2009) 106,
  [\href{http://xxx.lanl.gov/abs/0903.4665}{{\tt arXiv:0903.4665}}].

\bibitem{Frixione:1995ms}
S.~Frixione, Z.~Kunszt, and A.~Signer, {\it {Three jet cross-sections to
  next-to-leading order}},  {\em Nucl.Phys.} {\bf B467} (1996) 399--442,
  [\href{http://xxx.lanl.gov/abs/hep-ph/9512328}{{\tt hep-ph/9512328}}].

\bibitem{Campbell:2010ff}
J.~M. Campbell and R.~Ellis, {\it {MCFM for the Tevatron and the LHC}},  {\em
  Nucl.Phys.Proc.Suppl.} {\bf 205-206} (2010) 10--15,
  [\href{http://xxx.lanl.gov/abs/1007.3492}{{\tt arXiv:1007.3492}}].

\bibitem{mcfmweb}
J.~M. Campbell, R.~Ellis and C.~Williams,
``{MCFM web page}.'' \url{http://mcfm.fnal.gov}.

\bibitem{Whalley:2005nh}
M.~Whalley, D.~Bourilkov, and R.~Group, {\it {The Les Houches accord PDFs
  (LHAPDF) and LHAGLUE}},  \href{http://xxx.lanl.gov/abs/hep-ph/0508110}{{\tt
  hep-ph/0508110}}.

\bibitem{Nakamura:2010zzi}
{\bf Particle Data Group} Collaboration, K.~Nakamura et~al., {\it {Review of
  particle physics}},  {\em J.Phys.} {\bf G37} (2010) 075021.

\bibitem{LatundeDada:2006gx}
O.~Latunde-Dada, S.~Gieseke, and B.~Webber, {\it {A Positive-Weight
  Next-to-Leading-Order Monte Carlo for e+ e- Annihilation to Hadrons}},  {\em
  JHEP} {\bf 0702} (2007) 051,
  [\href{http://xxx.lanl.gov/abs/hep-ph/0612281}{{\tt hep-ph/0612281}}].

\bibitem{Cacciari:2011ma}
M.~Cacciari, G.~P. Salam, and G.~Soyez, {\it {FastJet user manual}},  {\em
  Eur.Phys.J.} {\bf C72} (2012) 1896,
  [\href{http://xxx.lanl.gov/abs/1111.6097}{{\tt arXiv:1111.6097}}].

\end{thebibliography}
\providecommand{\href}[2]{#2}\begingroup\raggedright\endgroup

\end{document}